\documentclass[useAMS,usenatbib,twocolumn]{mn2e}
\usepackage[]{graphicx}
\usepackage{url}
\newcommand\apjl{ApJ}%
\newcommand\apjs{ApJS}%
\newcommand\ao{Appl.~Opt.}%
\newcommand\aap{A\&A}%
\newcommand\ngrb{493 }

\title{Calibration of X-ray absorption in our Galaxy}
\author[R. Willingale et al.]
{R. Willingale\thanks{E-mail: rw@star.le.ac.uk},
R.L.C. Starling, A.P. Beardmore, N.R. Tanvir and P.T. O'Brien \\
University of Leicester, Department of Physics and Astronomy,
University Road, LE1 7RH, UK}

\begin{document}
\date{Accepted  Received ; in original form }

\pagerange{\pageref{firstpage}--\pageref{lastpage}} \pubyear{2010}

\maketitle
\label{firstpage}

\begin{abstract}
Prediction of the soft X-ray absorption along lines of sight through our Galaxy
is crucial for understanding the spectra of extragalactic sources, but
requires a good estimate of the foreground column density of 
photoelectric absorbing species.
Assuming uniform elemental abundances this reduces to having a good
estimate of the total hydrogen column density, $N_{Htot}=N_{HI}+2N_{H_{2}}$.
The atomic component, $N_{HI}$, is reliably provided using the mapped 21 cm
radio emission but estimating the molecular
hydrogen column density, $N_{H_{2}}$, expected for any particular direction,
is difficult.
The X-ray afterglows of GRBs are ideal sources to probe X-ray
absorption in our Galaxy because they are extragalactic, numerous, bright,
have simple spectra and occur randomly across the entire sky.
We describe an empirical method, utilizing \ngrb afterglows detected by
the {\em Swift} XRT, to determine $N_{Htot}$ through the Milky Way
which provides an improved estimate of the X-ray
absorption in our Galaxy and thereby leads to more
reliable measurements of the intrinsic X-ray absorption and, potentially,
other spectral parameters, for extragalactic X-ray sources.
We derive a simple function, dependent on the product of the
atomic hydrogen column density, $N_{HI}$, and
dust extinction, $E(B-V)$, which describes the variation of the molecular hydrogen column density, $N_{H_{2}}$, of our Galaxy, over the sky.
Using the resulting $N_{Htot}$ we
show that the dust-to-hydrogen ratio is correlated with the carbon
monoxide emission and use this ratio to estimate the 
fraction of material which forms interstellar dust grains.
Our resulting recipe represents a significant revision in Galactic
absorption compared to previous standard methods, particularly at low
Galactic latitudes.
\end{abstract}

\section{Introduction}
\label{s1}
Understanding the soft X-ray spectra of extragalactic sources
requires an estimate of the foreground photo-electric absorption due to gas
and dust in the Milky Way.
The total photo-ionization cross-section which gives rise to X-ray absorption in the ISM can be written
as the sum of contributions from 3 phases
\begin{equation}
 \sigma_{ISM}=\sigma_{gas}+\sigma_{molecules}+\sigma_{dust}.
\end{equation}
It is conventional to normalise the cross-section with respect to the total hydrogen column density, $N_{Htot}$
cm$^{-2}$ (molecular, neutral atomic and ionized), such that the X-ray spectrum observed at energy $E$ is
\begin{equation}
I(E)=\exp(-\sigma_{ISM}(E)N_{Htot}) I_{0}(E)
\end{equation}
where $I_{0}(E)$ is the source spectrum. Absorption in the soft band, 0.1-2 keV, is included in spectral
fitting using a detailed model of the cross-section, $\sigma_{ISM}$, and the hydrogen column density,
$N_{Htot}$, is usually a fitted parameter. The cross-sections for each of the phases are obtained by summing the photo-ionization cross-sections of the individual constituent atoms and
ions weighting their contributions by the abundances. Following the notation in \cite{1977A&A....61..339R}
the cross-section for the gas phase can be written as
\begin{equation}
\sigma_{gas}=\sum_{Z,i}A_{Z} \times a_{Z,i}\times (1-\beta_{Z,i})\times \sigma_{Z,i},
\label{eqgas}
\end{equation}
where the relative
abundance of element $Z$ with respect to hydrogen is $A_{Z}=N_{Z}/N_{Htot}$ and the fraction
of ions of this element in the i$^{th}$ ionization state is $a_{Z,i}=N_{Z,i}/N_{Z}$ and $\sigma_{Z,i}$ is the
total photo-ionization cross-section of this ionization state. The fraction of ions in the
gas phase is $1-\beta_{Z,i}$ so $\beta_{Z,i}$ is the fraction in other phases (usually dominated
by the fraction in dust grains).
Atoms deep within large grains are
shielded by the outer layers and therefore their contribution to the total
cross-section is reduced.
Similarly, some atomic absorption edges are modified if the atoms are incorporated into molecules
and the molecular cross-section is not a simple sum of the cross-sections of the constituent atoms/ions.
If the ISM is warm then higher ionization states are excited and the $a_{Z,i}$ values will be modified.
 But all these factors only introduce small changes and what really dominates the total cross-section are
the abundances of the atomic types, $A_{Z}$. It is usually assumed that these abundances have fixed values
across the Galaxy and so if we fix the value of $N_{Htot}$ the column densities of all other
atomic types can be determined and the expected cross-sections can be estimated. Apportioning these atoms between atomic gas, partially ionized gas, molecules and dust will normally introduce only minor 
perturbations to the total cross-section. Further details about this model of X-ray
absorption in the ISM are given by \cite{2000ApJ...542..914W}.

For extragalactic X-ray sources we can split the total measured absorption into two 
components including both the column density in our Galaxy, $N_{H,g}$, and the
excess column density in the host galaxy or elsewhere,
$N_{H,i}$. If we know the redshift of the host then we can 
express the excess column in the rest frame of the host. However, if the spectral resolution is
 modest then  individual absorption line features will not be visible and the energy profile of
absorption is not expected to be a strong function of redshift meaning
we cannot fit both components
simultaneously. We must fix $N_{H,g}$ if we want to extract a meaningful estimate of
$N_{H,i}$. Therefore fitted values of the excess absorption for extragalactic sources are critically dependent on our understanding of $\sigma_{ISM}$ and $N_{H,g}$ for different lines of sight through our Galaxy.


The dominant fraction of the total hydrogen column density is atomic,
$N_{HI}$, \citep[approximately 80\% on average][]{2000ApJ...542..914W}
and this is readily estimated using 21 cm radio emission
maps (e.g. the Leiden-Argentine-Bonn (LAB) survey \cite{2005A&A...440..775K}).
However,
mapping the distribution of molecular hydrogen throughout our Galaxy is difficult. The molecule
has no permanent dipole which makes any radiative transitions weak and hard to detect.
Surveys in the near-infrared using the $H_{2}$ line emission at 2.122 $\mu$m (like {\em UWISH2} , \cite{2011MNRAS.413..480F}) readily detect dense ($n_{H_{2}}>10^{3}$ cm$^{-3}$)
giant molecular clouds and supernova remnants
but fail to detect diffuse emission from the low density interstellar medium (ISM)
($n_{H}\sim1$ cm$^{-3}$) because of the surface brightness limit, $\sim10^{-19}$ W m$^{-2}$ arcsec$^{-2}$, and the presence of numerous stars.
Molecular hydrogen in the ISM is most readily observed using absorption lines
from the first six rotational levels (J = 0-5) seen in the ultraviolet between 900 and 1130 \AA.
A sparse map of the column density, $N_{H_{2}}$ molecules cm$^{-2}$,
 across the Galaxy has been produced using  {\em FUSE} 
UV spectroscopy  on 73 extragalactic targets,  \citep{2006ApJS..163..282W}, but with so few lines
of sight it is difficult to construct a reliable and detailed picture. Furthermore all the targets considered
were at high
Galactic latitude with a maximum total hydrogen column density of $\sim10^{21}$ cm$^{-2}$ so
the distribution at high density in or near to the Galactic plane is not included. These data indicate that
the molecular fraction
of the column density along the line of sight, defined as
\begin{equation}
f(H_{2})=\frac{2N_{H_{2}}}{N_{HI}+2N_{H_{2}}},
\label{eqfh2}
\end{equation}
is highly variable at high Galactic latitudes and is not tightly correlated with the atomic gas column density, $N_{HI}$ atoms cm$^{-2}$, or the total hydrogen column.

The afterglow emission of Gamma Ray Bursts (GRBs) is seen at all wavelengths from X-ray through to
radio and is thought to be synchrotron radiation arising from the shock produced when the GRB jet
impacts the surrounding medium (e.g. \cite{1992MNRAS.257P..29M,1998ApJ...497L..17S}).
The X-ray emission (0.3-10 keV) from GRB afterglows is routinely measured today
by the
 {\em Swift} satellite \citep{2004ApJ...611.1005G} and in almost all cases the soft X-ray spectrum is well
modelled by a simple power law continuum modulated by
 photoelectric absorption at low energies.  Absorption is expected from gas in the Milky Way
along the line of sight but for most {\em Swift} GRBs there is also evidence for significant absorption
in excess of the Galactic foreground. This is attributed to gas in the host galaxy and/or intergalactic clouds
along the line of sight and has been studied in detail by many authors, see
\cite{2011ApJ...734...26B,2012MNRAS.421.1697C,2011A&A...525A.113S,
2011A&A...533A..16W,2011ApJ...735....2Z} and references therein.
Because the X-ray afterglows are bright and have simple continuum spectra
they provide an excellent opportunity to study the cold absorbing material in our Galaxy as well
as the ISM in the distant host galaxies and this is what we turn attention to
here.

\section{Fitting GRB X-ray afterglow spectra}
\label{strial}
\begin{figure}\begin{center}
\includegraphics[height=10cm,angle=0]{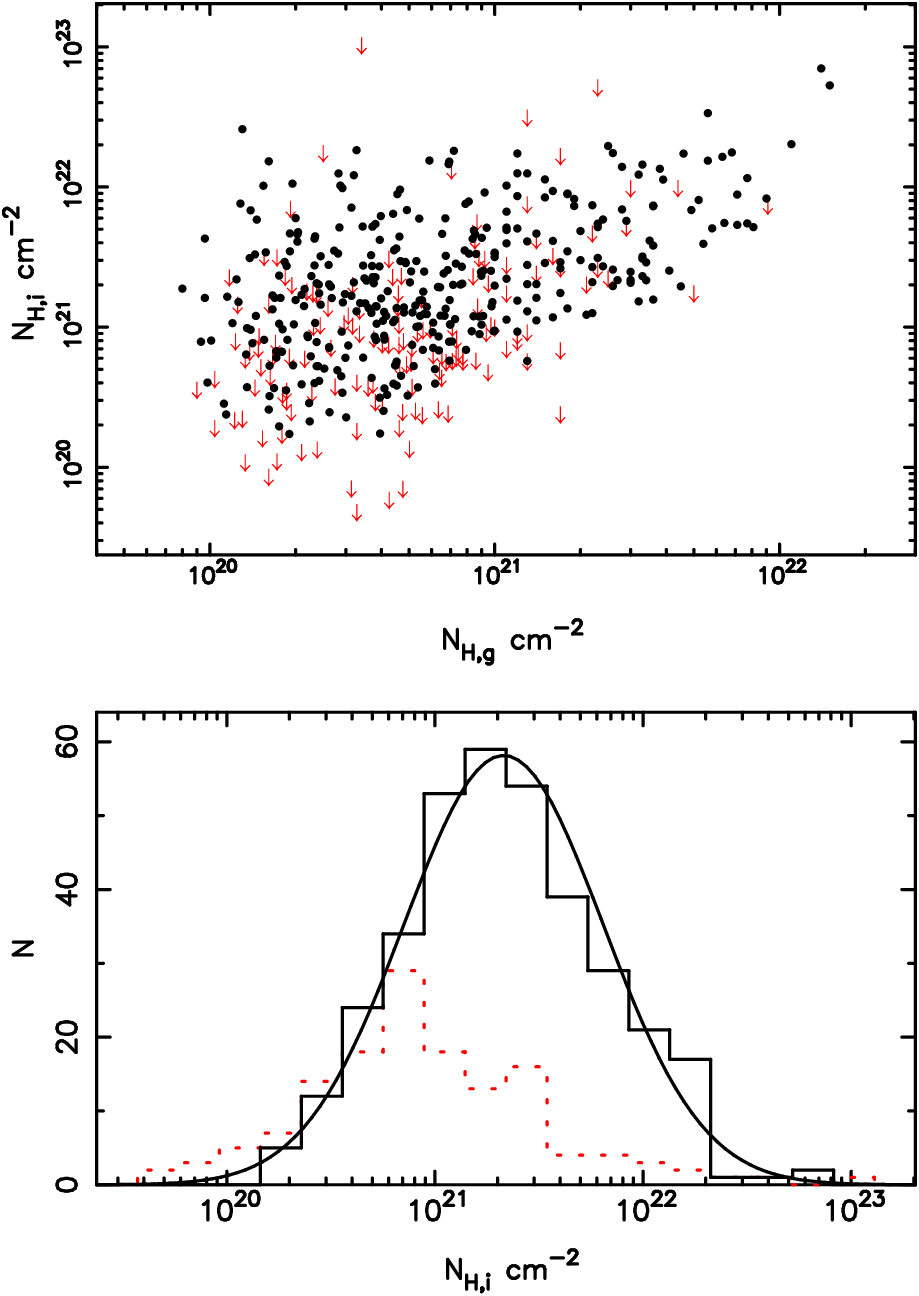}
\end{center}\caption{The results from fitting \ngrb individual GRB afterglow spectra. Top panel:
The fitted values of excess hydrogen column $N_{H,i}$ (assuming $z=0$) plotted
against the fixed values of $N_{H,g}$.
Upper limits (3 sigma) for $N_{H,i}$ are shown as arrows.
No correlation is expected between
these two column densities but even allowing for the upper limits
it is clear that $N_{H,i}$ is systematically increasing with $N_{H,g}$.
We note that a Galactic column density of $N_{H,g}=10^{21}$ cm$^{-2}$
corresponds to a average Galactic latitude of -13 degrees in the South
and +22 degrees in the North. Some 29\% of the sky has
$N_{H,g}>10^{21}$ cm$^{-2}$.
Bottom panel: The distribution of fitted $N_{H,i}$.
The smooth curve is a log-normal fit to the distribution.
The dotted histogram shows the distribution of the
upper limits.}
\label{fig1}
\end{figure}

We selected X-ray afterglow spectra measured by the {\em Swift}
X-ray Telescope (XRT) in photon counting (PC) mode from GRBs
up until 2011 November 3. We did not use data obtained in Windowed Timing (WT) mode to avoid,
 as far as possible, early times post-burst when spectral evolution, probably associated with the prompt
emission, is most likely to be occurring. For a few
GRBs there is evidence for an additional, early, thermal continuum component which is 
probably associated with the prompt emission or a supervova (SNe)
rather than the afterglow. One such object for
which extensive PC mode data were obtained
is GRB 060218 \citep{2006Natur.442.1008C} and we excluded this from our sample.
GRB 090618 \citep{2011MNRAS.416.2078P},
GRB 100316D \citep{2011MNRAS.411.2792S}
and GRB 101219B \citep{2012arXiv1207.1444S}
have similar early spectra but such complications are only
seen in WT mode so these were included in our analysis.
We fitted the GRB afterglow spectra with a simple model comprising
a power-law continuum specified by two fitted parameters, a photon index
and a normalisation, and two absorption components representing the
Galactic absorbing column (with fixed column density) and the excess
absorbing column for which the column density was allowed to float.
The $N_{H,g}$ values used were taken from the LAB 21 cm all-sky survey, \cite{2005A&A...440..775K}, using the FTOOLS NH  procedure on the GRB positions.
Fig. \ref{fig1} shows the values for the fixed
Galactic, $N_{H,g}$, and fitted excess $N_{H,i}$, column
densities from spectral fits of \ngrb afterglows.

The fitting was done using XSPEC version 12.7.0 \citep{1996ASPC..101...17A} utilizing the absorption model tbabs
\footnote{The XSPEC command variants ztbabs and tbvarabs
were also used extensively in the analysis to provide direct access to the
abundance, dust grain and redshift parameters.}, \cite{2000ApJ...542..914W}.
This model includes elemental abundances tabulated by the same authors and
these abundances 
were used for all the spectral fitting reported below except for
those cases mentioned explicitly in the text.
The $N_{H,i}$ values plotted are the
fitted values (excess absorption over and above the fixed Galactic value) assuming a redshift of zero
(even for those bursts for which we know the redshift). We note that 
in this paper we are not concerned with the physical origin of the excess
absorption in GRB afterglow spectra, and using the complete set,
irrespective of whether a redshift has been
determined, provides a much larger sample for the analysis and removes
any risk of optical bias.
The underlying GRB redshift distribution should, of course, be
independent of position with respect to the Galaxy.
The paucity of points in the bottom right of the plot is
not immediately surprising  because if the Galactic column is high
the sensitivity to an excess column is reduced. 
We would expect the points from the lower-right (low $N_{H,i}$
and high $N_{H,g}$) to
become upper limits in the upper-right of the distribution.
In fact there are very few upper limits for objects with high
$N_{H,g}$ and for these values the fitted
$N_{H,i}$ are higher than average.
The distribution of fitted $N_{H,i}$ values is shown in the
bottom panel.
The observed distribution is slightly skewed with respect to the log-normal fit and there appears to be an excess bump in the high tail.
The distribution has a mean value of $2.1\times10^{21}$
cm$^{-2}$ and a rms width of 0.48 dex.

Although the X-ray afterglows of GRBs are exceptionally bright compared with other extragalactic sources,
they are only visible over a relatively short time window and
the spectra of some bursts have limited statistics, which
of course is ultimately the reason for the
upper limits on excess absorption for many GRBs.
The distribution of upper limits is shown relative to the
whole distribution in the lower panel of Fig. \ref{fig1}.
They are clearly biased
towards the lower tail of the $N_{H,i}$ distribution, as expected,
and we need some
way of including these lower limits in our analysis if we are to
understand what is going on.
In order to investigate further we sorted the
bursts into ascending order of $N_{H,g}$ and selected GRBs in groups of 26, each group containing bursts that are expected to suffer a similar degree of  Galactic absorption. This gave 18 groups of 26 and a final 19th group of 25. We summed together the count spectra in each group to produce 19 composite spectra and then 
fitted each composite spectrum using XSPEC. Tools are provided 
along with XSPEC to sum the source spectra, background files,
the area files (ARFs) and the response files (RMFs) to facilitate accurate fitting of composite spectra.
The component spectra from the GRBs were extracted using grade 0-12 events
from PC mode and the appropriate ARF and RMF files were automatically
selected for the correct epoch. We didn't apply any binning to the raw
spectra because the composite spectra contained more than adequate counts.
Pile-up was avoided by excising the core of the PSF when appropriate.
The ARF files were summed weighting by the total counts in the
spectrum. It was found that only two significantly different RMF files
were required,
before and after the substrate voltage change which was performed 2007-09-01.
We summed these weighting by the accumulated time before and after
the change over date. We checked this summing procedure using different
weightings and different combinations of ARFs and RMFs. Because each
group contains 26 GRBs the results were remarkably
insensitive to the weightings or components used.
The average photon index of the power law continuum fitted for the individual
GRBs was $\Gamma=1.96$ and the rms scatter was small, 0.18. We therefore
expect the continua of the composite spectra to be well represented
by a simple power law (with similar photon index) and we
fitted the composite spectra using the same model as used for the
individual GRBs.
We fixed the redshift for the 
excess column as $z=0$ and fixed the $N_{H,g}$ to the mean for each
group.  Because the range of $N_{H,g}$ within each group is narrow
small changes in the fixed $N_{H,g}$ values
used make no difference to the results.
The mean value of the excess absorbing column density required increases with redshift but, crucially, the quality of the fit and the scatter of the excess absorption column
densities remain the same.

\begin{figure}\begin{center}
\includegraphics[height=8cm,angle=-90]{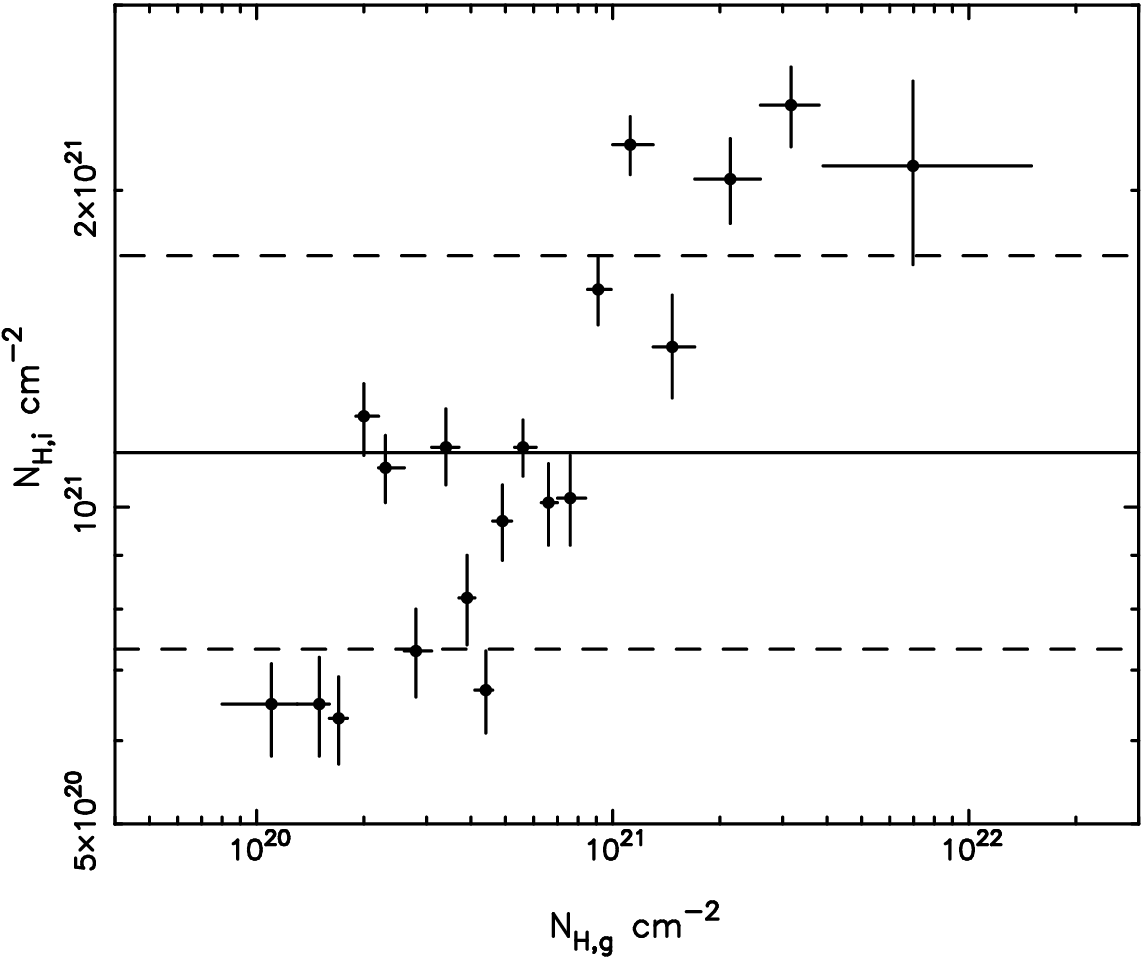}
\end{center}\caption{The fitted values of $N_{H,i}$, $z=0$, for 19 GRB groups plotted against the fixed values
 of Galactic hydrogen column $N_{H,g}$.
The solid horizontal line indicates the geometric mean
and the dashed lines $\pm1$ sigma in dex.}
\label{fig2}
\end{figure}

Fig. \ref{fig2} shows the fitted $N_{H,i}$ values for the 19 groups plotted against the fixed $N_{H,g}$
values. Following the conventional approach, already used for the fits shown in Fig. \ref{fig1}, we used the
$N_{H,g}$ values taken from the LAB 21 cm all-sky survey. 
The error bars in $N_{H,g}$ indicate the range of Galactic absorption within each group while those in
$N_{H,i}$ represent the 90\% range for the fitted values.
Crucially all the groups return a well defined $N_{H,i}$ value
with a 90\% range and there are no upper limits.
We expect there to be no correlation between
the excess absorption in the GRBs and the column density within our Galaxy but the result of fitting
the groups clearly indicates a strong trend.
We can quantify the significance
of the trend by calculating a reduced Chi-squared value against a constant
model for $N_{H,i}$, $\chi^{2}_{\nu}=64.4$. 
In evaluating $\chi^{2}$ we use the formal fitting errors, although this
will underestimate the true uncertainties given the random variation in the
GRB afterglow properties from group to group.
Assuming the width of the distribution of $N_{H,i}$ values in Fig. \ref{fig1}
is representative of the GRB population, convolved with the measurement
errors, we expect the rms between groups of
26 GRBs to be $\approx0.09$ dex. Using this we can estimate the 
contribution to reduced Chi-squared from the scatter in $N_{H,i}$ giving
an expected value of reduced Chi-squared as $<\chi^{2}_{\nu}>=17.3$. 
The trend is obviously significant.
Using linear regression we estimate that the total swing in the $N_{H,i}$
values fitted across the full range of $N_{H,g}$ values in the groups
is a factor of 6.0.
Either there is something wrong with the calibration of the data,
the spectral fitting procedure is faulty or the model for the Galactic absorption is incorrect.  

There is a spread of GRB brightnesses within each group but if only
one or two GRBs were to dominate this might bias the results.
In fact the brightest GRB in each group contributes somewhere 
in the range 10-15\%
of the total count so a single GRB is not dominant in any group.
However the distribution of brightnesses of the GRBs in each group is
obviously important. It reduces the effective number of GRBs contributing
to the averaging and thereby increases the scatter we expect from
the redshift dependence and intrinsic spread in $N_{H,i}$ values.
We return to this point later and show that the chosen grouping of the
GRBs does not bias the results.

\begin{figure}\begin{center}
\includegraphics[height=8cm,angle=-90]{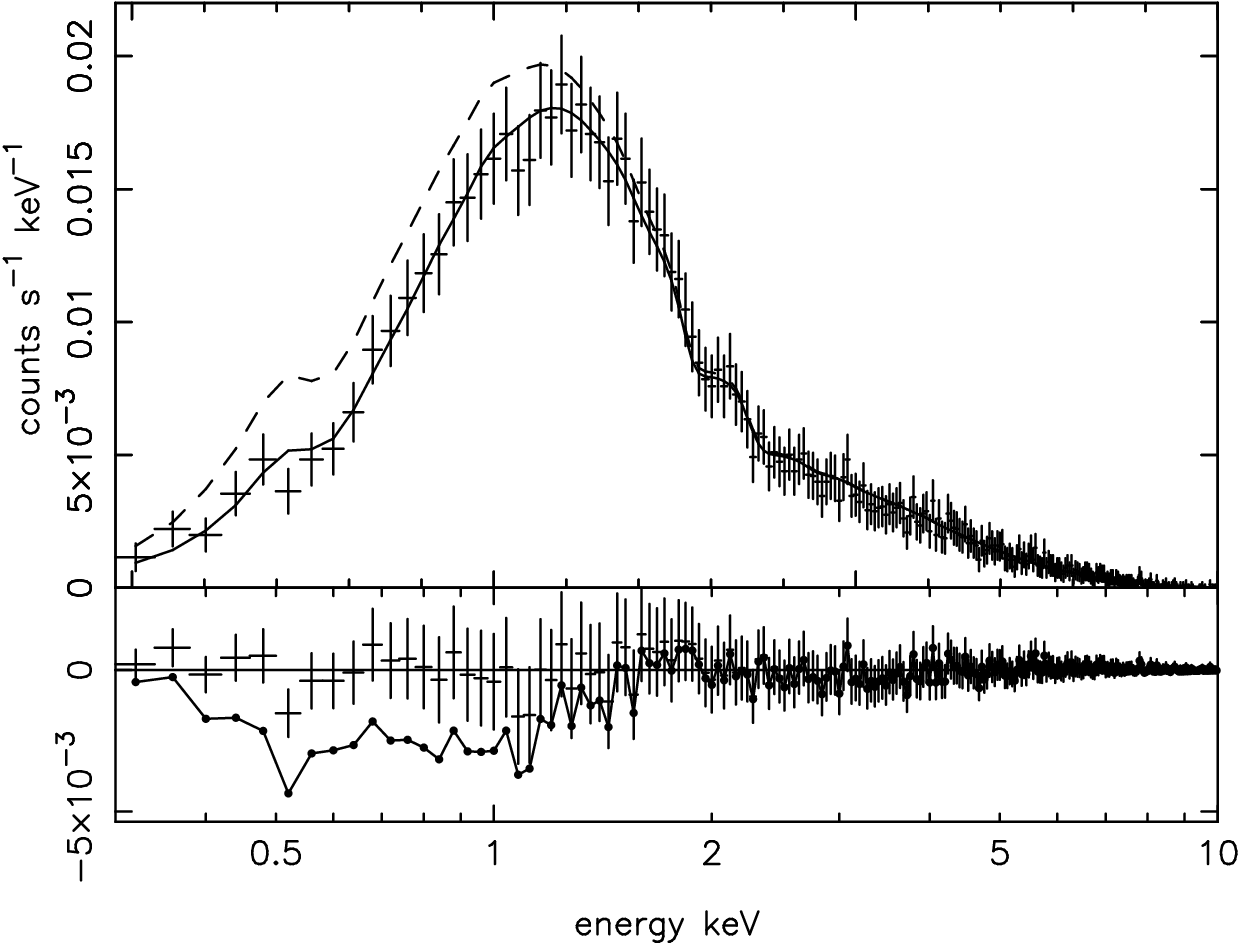}
\end{center}\caption{Top panel: The measured count spectrum for the 
16$^{th}$ composite group. The solid line indicates
 the best fit model. The dashed line shows the model prediction when $N_{H,i}=0$. Bottom panel: The
count rate residuals. The solid curve and dots show the residuals when $N_{H,i}=0$.}
\label{fig3}
\end{figure}

We inspected all the group fits to check for anomalies. Fig. \ref{fig3} shows the count rate spectrum and
fitted profile for the 16$^{th}$ group which is typical of the groups with $N_{H,g}>10^{21}$ cm$^{-2}$.
The solid curve is the best fit including the
fixed Galactic column, $N_{H,g}$ and the fitted excess column, $N_{H,i}$. The statistics are excellent and there are no features or large residuals to suggest
a fault with the calibration or the fitting procedure. The dashed line is the predicted spectrum with
the excess absorption component set to zero, $N_{H,i}=0$. The count rate residuals are shown in the lower
panel. The excess absorption signature
is clearly significant and covers a wide range of the low energy bins. For the GRB groups with
$N_{H,g}<10^{21}$ cm$^{-2}$ the excess absorption signature is much larger than the example
shown in Fig. \ref{fig3}.

The XRT calibration influences the fitting of the absorption component
in three ways. The first and dominant factor is the roll-off of the
efficiency below $\sim1.0$ keV. If the efficiency roll-off is too shallow then
the fitted column density must increase to match the detected count rate
profile. Conversely, if the roll-off is too steep, the fitted column
density will be too low. The second factor is the broad efficiency profile
and, in particular, the roll-off of the efficiency above $\sim2$ keV and
the ratio of the collecting area at high energies, above $\sim2$ keV,
to that at low energies, below  $\sim1$ keV. These factors determine the
profile of the power law continuum spectrum. If the high energy
roll-off is too shallow then the fitted spectral index will be
too large and the fitted column density must increase to compensate.
Conversely, if the fitted index is too small then the fitted column
density will be too small.
The third factor is the redistribution
matrix which dictates the detailed shape of the predicted count
distribution. Any mismatch between the predicted and observed count
profile degrades the goodness of fit and can lead to systematic
errors in the fitted parameters including the column density.

\begin{figure}\begin{center}
\includegraphics[height=8cm,angle=-90]{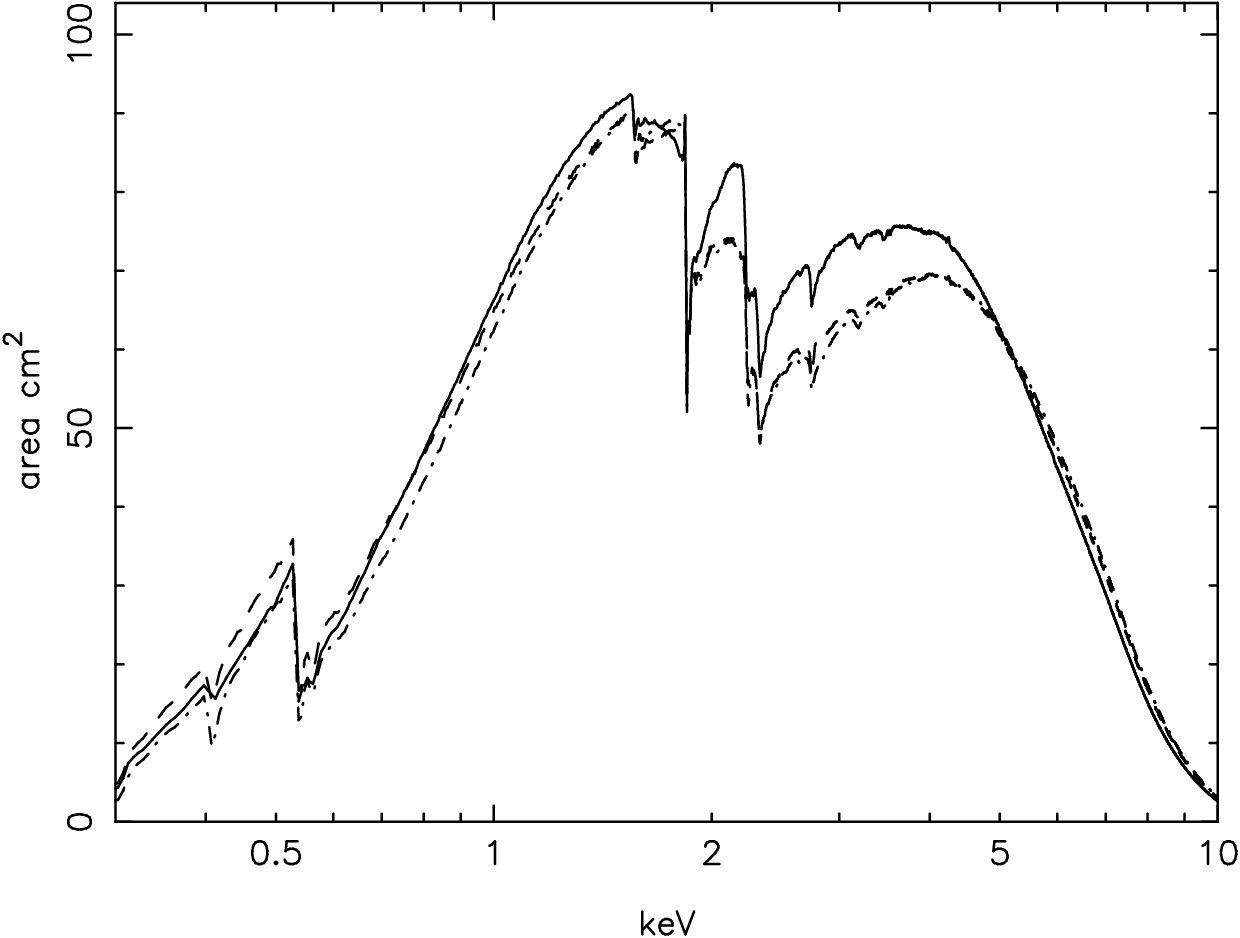}
\end{center}\caption{Comparison of the
effective area as a function of energy for the current (version 16)
XRT calibration (solid curve),
our trial1 calibration (dashed line), and our trial2 calibration
(dot-dashed line). The trial1 and trial2 values have been scaled so
all three curves have the same area at $\sim1.85$ keV.}
\label{fig4}
\end{figure}

According to the latest XRT calibration release note (version 16)
a systematic error  of $\sim3\%$ is required when fitting
high statistical quality spectra using the current RMF and ARF response files. Including such a systematic
error has no effect on the results plotted in Fig. \ref{fig2}.
The magnitude of the absorption fitted depends on the area calibration as
a function of energy, in particular over the low energy range 0.3-1.5 keV.
If the roll-off of this area set by the calibration 
is incorrect, the absorption will be over- or under-estimated accordingly
as discussed above. In order to determine 
the sensitivity of the fitted $N_{H,i}$ results reported throughout this work we repeated all the
analysis with three versions of the {\em Swift} XRT calibration. Firstly the currently released
version (SWIFT-XRT-CALDB-09\_v16) and
secondly two trial versions generated by 
employing plausible enhancements of the known uncertainties
in the hardware calibration:
the detector QE, the filter transmission using
updated absorption coefficients and measurements and mirror X-ray reflectivity
incorporating a contaminating carbon overcoat.
The trial calibrations comprise both new RMF and ARF files. A comparison of the
areas for these calibrations
is shown in Fig. \ref{fig4}.
Significant differences are the 
depth of the oxygen K-edge at $\sim0.54$ keV and the detailed structure across
the silicon K-edge at $\sim1.8$ keV and the gold M-edges $2.2-4.0$ keV.
The trial1 and trial2 curves in Fig. \ref{fig4} have been
scaled so that the differences in the roll-off at low energies is
most obvious. 
For trial1 the roll-off below 1 keV is $\sim20\%$ shallower than the current
calibration and for trial2 this roll-off is $\sim20\%$ steeper.
The differences between
the true calibration and either version 16, trial1 or  trial2
are very unlikely to
be greater than the differences visible in Fig. \ref{fig4}.
We note that the trial XRT calibrations are not intended as a
better approximation
to the true state of the instrument
but they serve to illustrate that significant changes to the effective area
calibration do not alter our results.
Using the trial1 XRT calibration increases the fitted $N_{H,i}$ values by
around 25\% and the total swing in the $N_{H,i}$
values fitted across the full range of $N_{H,g}$ values in the groups
is a factor of 5.7, a little less than the swing using the
current calibration (6.0).
Using the trial2 decreases the fitted $N_{H,i}$ by
around 20\% and increases the swing factor to 10.4.
somewhat larger than for the current calibration.
In addition to these extreme versions of the calibration
we also tried varying the RMFs and ARFs individually through 
intermediate values. These changed the results in the same sense as
trial1 or trial2 although the changes were always smaller.

The primary influence of all the calibration factors on the fitted column
density is a simple scaling because the instrument response and the fitting
procedure are, essentially, linear. If we change the roll-off in the
effective area below $\sim1$ keV by 10\% then we expect the fitted $N_{H}$ to
change by $\sim10$\%. This scaling is independent of the absolute value of
$N_{H}$. However, the trend in the fitted column density values which we are
trying to account for is not the result of a simple scaling error. 
We conclude that uncertainties in the calibration cannot account for
the observed trend seen in Fig. \ref{fig2},
and we believe that the XSPEC fitting  procedure is well tried and tested.

\section{Modifying the Galactic absorption model parameters}
\subsection{Global scaling}
\begin{figure}\begin{center}
\includegraphics[height=8cm,angle=-90]{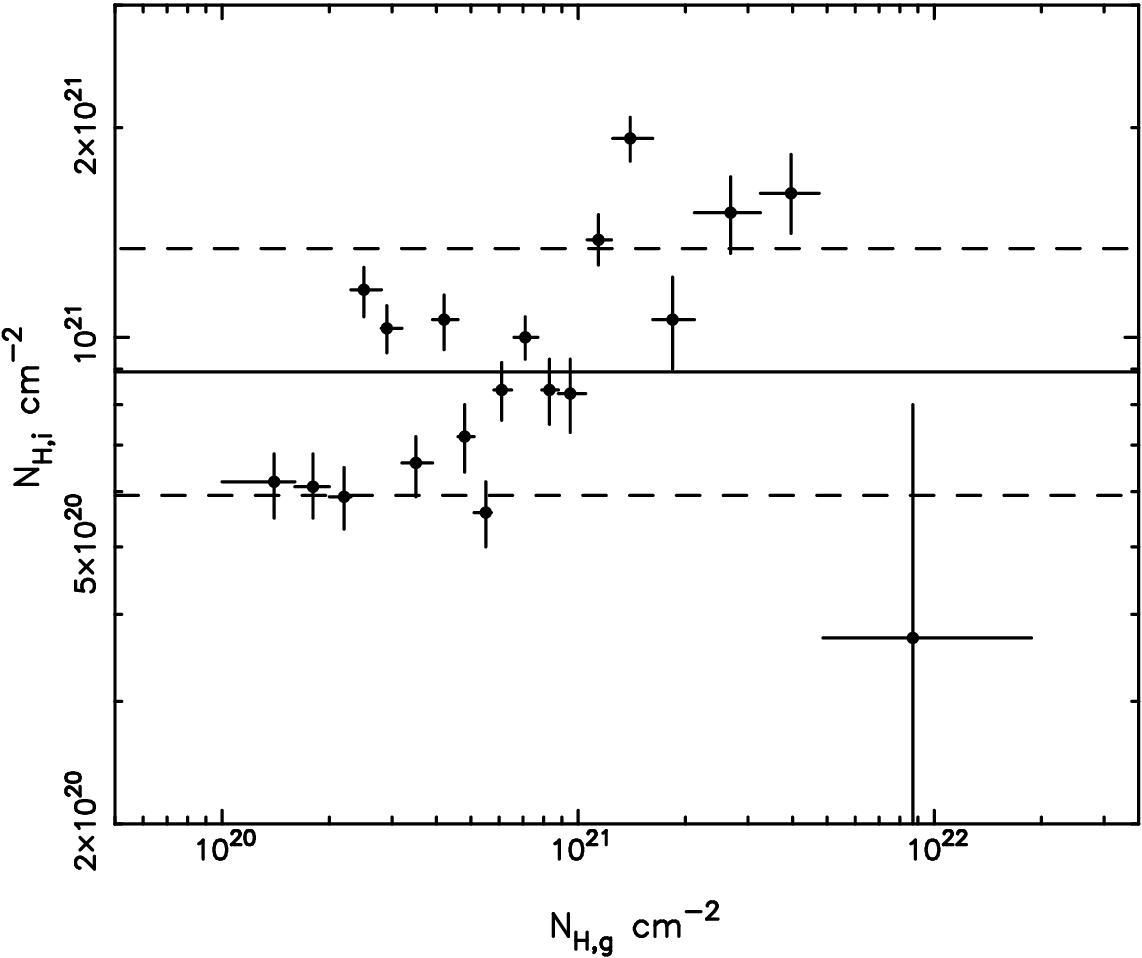}
\end{center}\caption{
The $N_{H,i}$ values for the 19 GRB groups using $N_{H,g}$ equal to $1.25\times N_{HI}$.
The apparent discrepancy between the bin to bin scatter
and the 90\% error bars may not be
very surprising, given that each group contains
GRBs at a range of redshifts and true (intrinsic) columns.
The horizontal lines are as in Fig. \ref{fig2}.}
\label{fig5}
\end{figure}

The default implementation of the \cite{2000ApJ...542..914W} ISM model in the XSPEC tbabs model 
includes a molecular hydrogen fraction, $f(H_{2})$ as defined in 
Equation \ref{eqfh2}, of 20\%. It assumes that the $N_{Htot}$ fit parameter is
the sum of the column densities (both expressed in terms of 
hydrogen atoms per unit area) of 
$HI$ and $H_{2}$. Therefore, employing the widely adopted
conventional approach and setting $N_{Htot}=N_{HI}$ from 21 cm surveys is strictly an incorrect use of
this routine. The measured $HI$ column should be multiplied by 1.25 so that 20\% represents the
molecular column and 80\% the $HI$ fraction. Fig. \ref{fig5} shows the distribution of
fitted $N_{H,i}$ if we adopt this value for the Galactic column in the group spectral fitting.
The bursts in the last (19th) group
have a very high mean foreground $N_{H,g}\sim10^{22}$ cm$^{-2}$,
and are typically at galactic latitudes $<2.0$ degrees.
Thus their lines of sight pass through
complex regions of the disk, and it would be rather surprising if
estimates of the foreground absorption were at all accurate.
Nonetheless, in Fig. \ref{fig5} the 19th
group is, within the large formal errors, consistent with the $N_{H,i}$
value for the low foreground bins.
The remaining groups still show the same unexplained trend as before.
The reduced Chi-squared is now $\chi^{2}_{\nu}=40.9$ while the expected value
is $<\chi^{2}_{\nu}>=13.5$.
If the trial1 calibration described in Section \ref{strial} is used
Fig. \ref{fig5} is very similar except
the mean value of $N_{H,i}$ is increased by $~25\%$.
The trial2 calibration decreases the mean value by $18\%$.

\begin{figure}\begin{center}
\includegraphics[height=8cm,angle=-90]{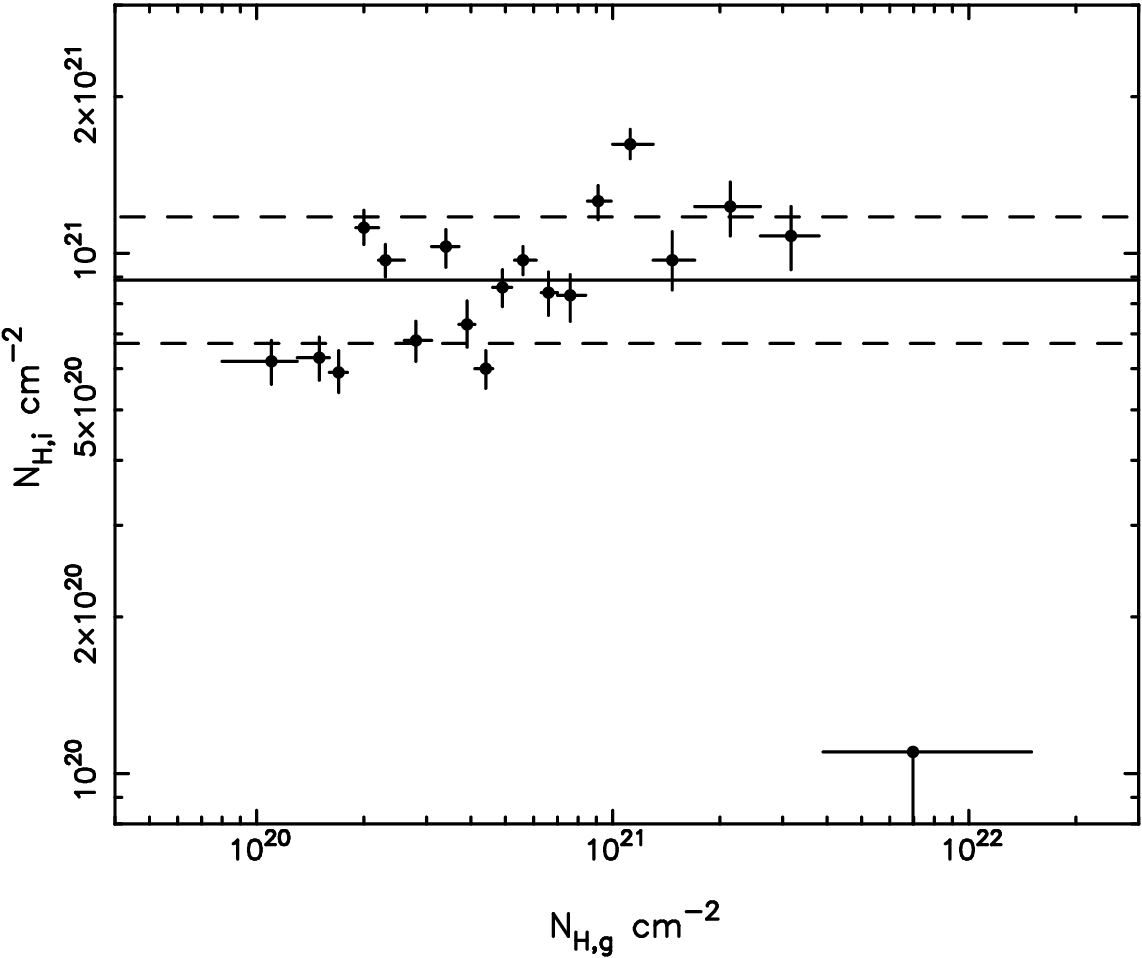}
\end{center}\caption{
The $N_{H,i}$ values for the 19 GRB groups obtained using the abundances of Anders \& Grevesse (1989).
The horizontal lines are as in Fig. \ref{fig2}.}
\label{fig6}
\end{figure}

A simple scaling of the measured $N_{HI}$ column density (or equivalently including a fixed percentage for
molecular hydrogen) is unable to explain the apparent variation in
excess GRB column density as a function of the absorbing column in our Galaxy. 
An alternative explanation for
the apparent variation might be an
error in the elemental abundances assumed. The \cite{2000ApJ...542..914W} model includes updated
abundances which are approximately 70\% of the previously widely used values from
\cite{1989GeCoA..53..197A}.
To test the effect of this change
we re-ran the group fitting using the \cite{1989GeCoA..53..197A}
abundances for the $N_{H,g}$ component. The result is shown in Fig. \ref{fig6}.
The apparent variation remains and the Galactic absorption used for the top $N_{H,g}$ group is now too
large and results in an upper limit which is lower than the fitted values for all the other groups.
Using the trial calibrations
to produce Fig. \ref{fig6} makes little difference.
Using trial1 the $N_{H,i}$ value for the 19$^{th}$ group is still too
low but somewhat higher and closer to the average value.
If the trial2 calibration is used
then the 19$^{th}$ is near the average value but $N_{H,i}$ for
the lower $N_{H,g}$ drop away in the same trend as Fig. \ref{fig5}.

 We also tried varying the global percentage of molecular hydrogen from 0 to 30\% and varying the
elemental abundances between values consistent with \cite{2000ApJ...542..914W} and 
\cite{1989GeCoA..53..197A}
but no combination improved the situation. We conclude that the problem is not resolved by a simple
scaling of either the hydrogen density or the elemental abundances independent of position on the sky.
What is required is a modification of the ISM model as a function of 
direction across the Galaxy. In particular the absorption associated with the mid-range values of $N_{H,g}$,  around $10^{21}$ cm$^{-2}$, needs to be increased while the lowest and the very highest absorption values
should remain as they are. In all the subsequent analysis we reverted to using the abundances given in \cite{2000ApJ...542..914W}.

\subsection{The column density of molecular hydrogen as a function of $N_{HI}$}
\begin{figure}\begin{center}
\includegraphics[height=10cm,angle=0]{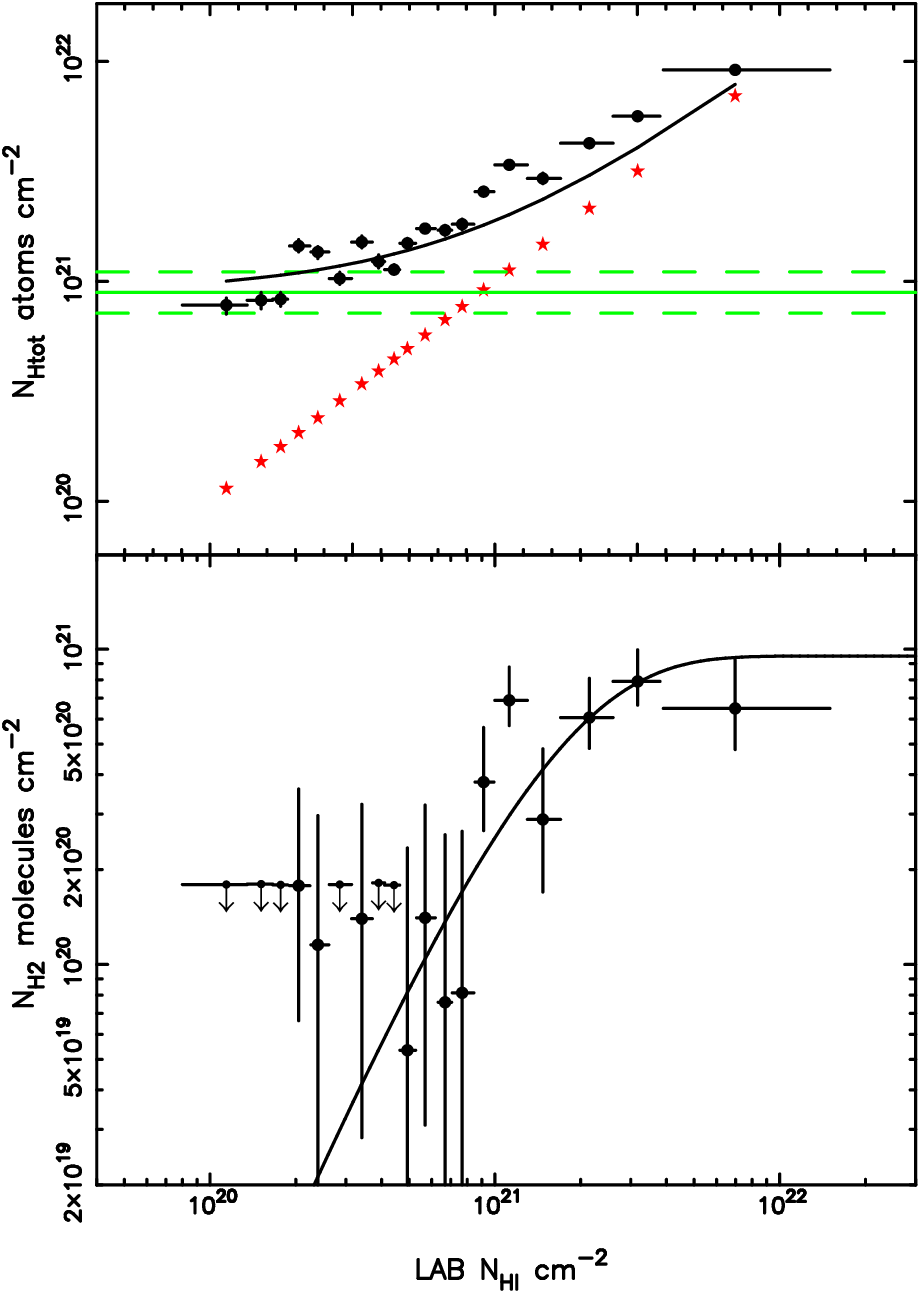}
\end{center}\caption{Upper panel: 
The fitted total hydrogen column densities, $N_{Htot}$, for the 19 group spectra
plotted vs. the $N_{HI}$ average for each group. The stars show the $N_{HI}$ component (from the LAB survey) and the
horizontal lines indicate the mean and rms range of the expected excess column density
component, $N_{H,i}$, constant across all groups. The solid rising curve is the sum of LAB $N_{HI}$ and $N_{H,i}$.
Lower panel: The $N_{H_{2}}$ column density derived by subtracting the Galactic HI and excess HI
contributions from the total. Those $N_{H_{2}}$ values plotted with downward
pointing arrows are upper limits.
The solid curve shows the simple model adopted for the molecular hydrogen
column density as a function of the HI column density.}
\label{fig7}
\end{figure}

The fraction of hydrogen in molecular form is known to be highly variable
\citep[e.g. ][]{2002ApJ...577..221R}.
In order to get a handle on the possible variation in the column density of molecular hydrogen
we fitted the 19 group spectra with a single absorption component giving a total
column density estimate, $N_{Htot}$, for each
group. This fitted total is expected to be the sum of the Galactic and
excess components.
Fig. \ref{fig7} shows these single fitted values as a function of the average LAB survey $N_{HI}$ values for each group.
We expect $N_{Htot}$ to be the sum of contributions from Galactic $N_{HI}$, excess $N_{H,i}$ and
Galactic $N_{H_{2}}$.
The lower panel shows the estimate of $N_{H_{2}}$ obtained by subtracting
the other two components from the total. 
To produce these values we assumed that the excess
component has
a mean of $9.4\times10^{20}$ cm$^{-2}$ (which is the average for the
first 4 groups and very close to $8.9\times10^{20}$ cm$^{-2}$ 
given by the results in Fig. \ref{fig5}) 
and an rms value
estimated from the width of the distribution from the individual fits
shown in Fig. \ref{fig1}.
For $N_{HI}>10^{21}$ cm$^{-2}$ the putative molecular hydrogen column is high and constant, independent of $N_{HI}$,  and below
this it drops away although it is difficult to determine the functionality of the
decline because the sensitivity is limited by the intrinsic scatter of the $N_{H,i}$ values for
the individual GRBs. Note that 6 of the groups below $N_{H,g}=5\times10^{20}$
cm$^{-2}$ have $N_{H_{2}}$ values which are upper limits.
We repeated the analysis summarised in Fig. \ref{fig7} using the trial1
calibration described in Section \ref{strial}.
The $N_{Htot}$ values returned were 
$2.2\times10^{20}$ cm$^{-2}$ higher, averaged across all groups.
For groups with $N_{Htot}<1.5\times10^{21}$
cm$^{-2}$ this corresponds to an average increase of $16\%$ while for
those with $N_{Htot}>1.5\times10^{21}$ the average increase is
$8\%$.
We had to increase the assumed
mean value of the excess component to $1.1\times10^{21}$ cm$^{-2}$,
in line with the trial1 calibration results returned
when plotting Fig. \ref{fig5}, but the estimated
$N_{H_2}$ values in the lower panel were the same within the limits imposed by
the error ranges plotted. Using the trial2 calibration shift things in 
the opposite direction with $N_{Htot}$ $\sim18\%$ down but otherwise the
result is the same.
This confirms that the excess Galactic
absorption, attributed to the 
molecular hydrogen column density in Fig. \ref{fig7},
is not an artifact of uncertainties in the calibration. It also 
demonstrates that the limits of uncertainties in the calibration introduce
a systematic error of $\sim16\%$ in the total fitted column density
for low columns and $\sim8\%$ when $N_{Htot}>1.5\times10^{21}$ cm$^{-2}$.

\begin{figure}\begin{center}
\includegraphics[height=8cm,angle=-90]{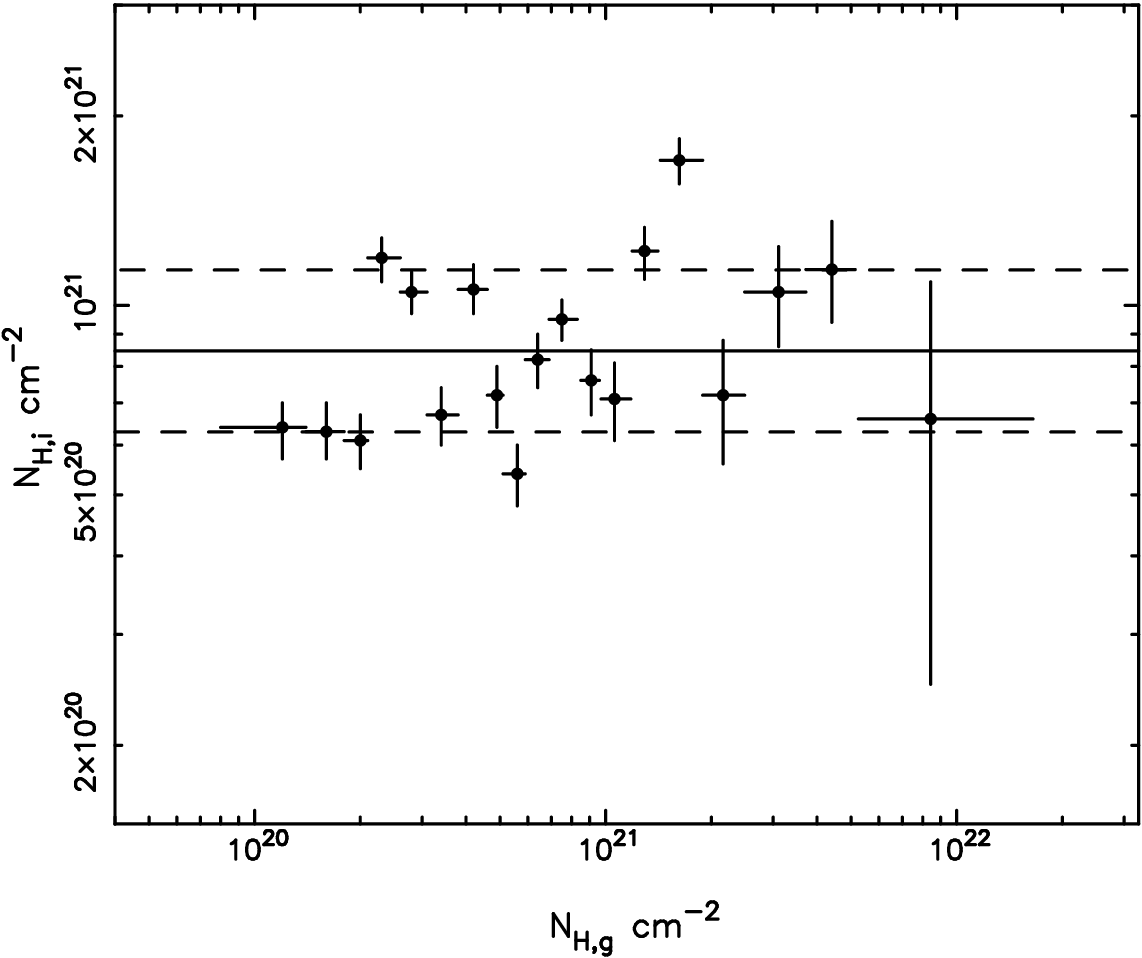}
\end{center}\caption{
The $N_{H,i}$ values for the 19 GRB groups using $N_{H,g}$ from 
Equation \ref{eqH} and using $N_{H_{2}}$ from Equation \ref{eqH1}.
The errors  plotted for $N_{H,i}$ are the 90\% ranges.
The horizontal lines are as in Fig. \ref{fig2}.}
\label{fig8}
\end{figure}

Given the results plotted in Fig. \ref{fig7} we initially assumed
that the molecular hydrogen column density throughout the
Galaxy has the profile indicated by the solid line plotted in lower
panel of Fig. \ref{fig7}. This has the form
\begin{equation}
N_{H_{2}}=N_{H_{2}max}\left[1-\exp\left(\frac{-N_{HI}}{N_{c}}\right)\right]^{\alpha}
\label{eqH1}
\end{equation}
and we set the total Galactic hydrogen column density as
\begin{equation}
N_{H,g}=N_{HI}+2N_{H_{2}}
\label{eqH}
\end{equation}
taking $N_{HI}$ from the LAB 21 cm survey. If $N_{HI}<<N_{c}$ then $N_{H_{2}}$ approximates to
a simple power law with index $\alpha$ and  if $\alpha=1$, $N_{H_{2}}$ is a simple fraction of $N_{HI}$.
If $N_{HI}>>N_{c}$ then $N_{H,g}$ asymptotes to $N_{H_{2}max}$. So $N_{c}$ is a characteristic
hydrogen column density where the powerlaw increase in $N_{H_{2}}$ flattens off to form a plateau with
the maximum value $N_{H_{2}max}$.
The profile in the lower panel of Fig. \ref{fig7}  was plotted using 
$N_{H_{2}max}=7.5\times10^{20}$ cm$^{-2}$,
$N_{c}=1.5\times10^{21}$ cm$^{-2}$ and $\alpha=2$.
We did not pursue any direct fitting of this profile using
the fitted $N_{Htot}$ values
because of the difficulty of handling the upper limits for $N_{H_{2}}$ that
arise when $N_{HI}$ is low.
Instead we performed a 3 dimensional grid search to find
the best fit model which gave the least rms scatter for $N_{H,i}$ across the
19 groups.
At each grid point we fitted the 19 group spectra with 2 
absorbing components, $N_{H,i}$ and a fixed $N_{H,g}$ including $N_{H_{2}}$
specified by values of the parameters $N_{H_{2}max}$, $N_{c}$ and $\alpha$.
Fig. \ref{fig8} shows the result with 
$N_{H_{2}max}= 7.5\times 10^{20}$ molecules cm$^{-2}$, $N_{c}=2.37\times 10^{21}$ atoms cm$^{-2}$
and $\alpha=2$.
The scatter in $N_{H,i}$ across the groups is now significantly reduced to
rms 0.128 dex closer to the expected value of $\approx0.09$ dex.
The reduced Chi-squared is $\chi^{2}_{\nu}=33.1$ while the expected value
given the intrinsic scatter in $N_{H,i}$ (rms 0.09 dex between the
groups of 26 GRBs) is $<\chi^{2}_{\nu}>=13.8$.
So some of the scatter 
in Fig. \ref{fig8} most likely arises from errors in modelling the Galactic column density.
The geometric mean value of $N_{H,i}$ for the 19 groups in Fig. \ref{fig8} is
$8.8\times10^{20}$ cm$^{-2}$ ($z=0.0$) which
is a significantly lower than the value for the distribution in Fig. \ref{fig1} ( $2.1\times10^{21}$ cm$^{-2}$)
and very close to the mean in Fig. \ref{fig5} and the
value assumed to plot Fig. \ref{fig7} ($8.9\times10^{20}$ cm$^{-2}$). This is
understandable since our estimate of the Galactic contribution to the absorption has now increased
and the upper limits present in the fitting of individual sources have been eliminated using the groups.
The estimated mean and rms scatter values of the excess component are plotted in Fig. \ref{fig7} as the
horizontal lines. They largely encompass
the fitted $N_{Htot}$ values for low $N_{HI}$ where the
excess column dominates. Finally, we note that the fitted values of
$N_{H_{2}max}$ and $N_{c}$ are dependent on the $N_{HI}$ values used.
Authors of the LAB survey and similar 21 cm surveys caution that
saturation of the 21 cm emission may lead to inaccurate or
under estimates of the atomic hydrogen column density when
$N_{HI}>>10^{21}$ cm$^{-2}$. Correlation of the LAB 21 cm map with
extinction maps (E(B-V) derived from IR emission, see next section)
indicates that any saturation is not very significant until
$N_{HI}>10^{22}$ cm$^{-2}$.
However, we acknowledge that some fraction of the
column we are ascribing to $H_{2}$ may actually be $HI$
that has been lost due to such saturation.

\subsection{Including dust extinction in the ISM modelling}
So far we have not considered interstellar dust.  Dust is relevant because it
locks up some metals from the ISM, which are not therefore seen in gas phase
abundance measurements.  As noted previously, these metals still contribute
to the soft X-ray absorption, albeit with some shielding due to the thickness
of the grains. We assume that this absorption is already correctly accounted
for (see also Section \ref{comparison}),
but dust has an additional role in catalysing molecule 
formation, and we consider that process in this section.
Most of the $H_{2}$ in the ISM is probably created on the surface of dust grains. The production rate is
dependant on
the collision probability between H atoms and the grains and the amount of gaseous molecular material 
released will depend on the sticking probability of $H_{2}$ on the grains. 
The competing dissociation rate depends on the
temperature, cosmic ray and X-ray flux and collisions within interstellar shocks. For details the reader is referred 
to the full
discussion in \cite{1982ARA&A..20..163S} and the references therein. For lines of sight out of the Galactic plane,
with cold, low density material, we expect the $N_{H_{2}}$ column density to be proportional to
the product $N_{HI}N_{dust}$. As we move closer to the plane the density will increase although
the ratio of $N_{H_{2}}/N_{HI}$ will be variable depending on local conditions. 
The profile derived from our initial analysis, shown in Fig. \ref{fig7}, indicates
that there is an upper limit to the local $N_{H_{2}}$ of $\sim10^{21}$  molecules cm$^{-2}$, independent of $N_{HI}$.

\begin{figure}\begin{center}
\includegraphics[height=8cm,angle=-90]{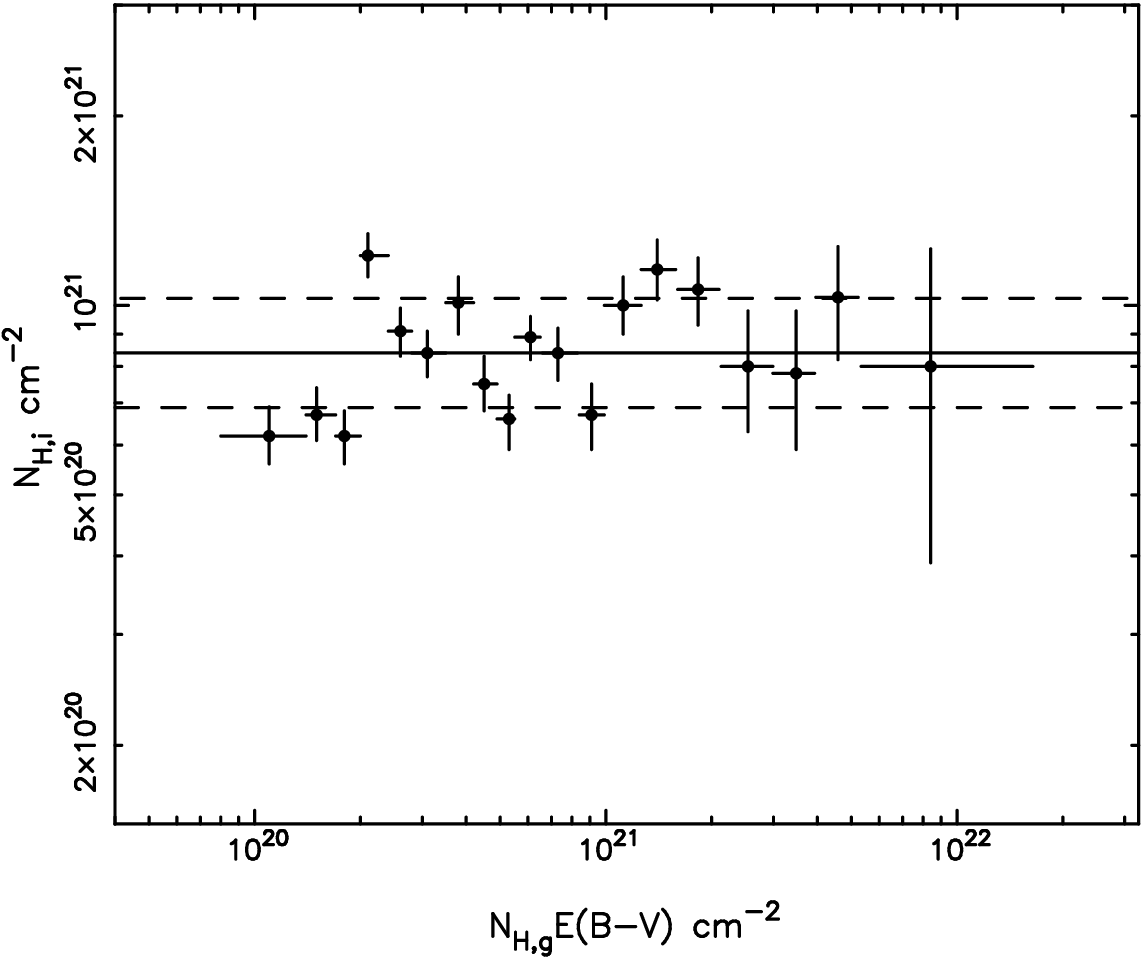}
\end{center}\caption{
The $N_{H,i}$ values for the 19 GRB groups
using $N_{H,g}$ from 
Equation \ref{eqH} with $N_{H_{2}}$ from Equation \ref{eqH2}.
The errors plotted for $N_{H,i}$ are the 90\% ranges.
The horizontal lines are as in Fig. \ref{fig2}.}
\label{fig9}
\end{figure}

We modified the functionality of the $N_{H_{2}}$ profile so that it includes
the extinction measure $E(B-V)$, using
the all-sky maps of $E(B-V)$ 
produced by \cite{1998ApJ...500..525S} from {\em IRAS} and
{\em COBE/DIRBE} infra-red 100 $\mu$m and 240 $\mu$m data.
These $E(B-V)$ values are derived by correcting the measured 
IR emissivity using the
measured temperature and assuming a constant reddening law so that $E(B-V)$ is
directly proportional to the dust column density, $N_{dust}$.
In order to produce
$N_{HI}$ and $E(B-V)$ values with the same limited angular resolution we rebinned the extinction maps
into an all-sky Aitoff projection array with a pixel size of 0.75 degrees. This is identical to the
array used by the FTOOLS NH procedure so we could use this to return both the LAB survey
$N_{HI}$ and $E(B-V)$ values for every GRB position (or any other position). The molecular
hydrogen column density profile was set as
\begin{equation}
N_{H_{2}}=N_{H_{2}max}\left[1-\exp\left(\frac{-N_{HI}E(B-V)}{N_{c}}\right)\right]^{\alpha}.
\label{eqH2}
\end{equation}
Fitting the grouped spectra adopting this measure for $N_{H_{2}}$ and, again, searching for
values of the parameters $N_{H_{2}max}$, $N_{c}$ and $\alpha$ which produced the minimum rms dex
scatter gave the results shown in Fig. \ref{fig9}.
The best fit parameters are $N_{H_{2}max}=(7.2\pm0.3)\times10^{20}$ molecules cm$^{-2}$,
 $N_{c}=(3.0\pm0.3)\times 10^{20}$ atoms cm$^{-2}$ and
 $\alpha=1.1\pm0.1$ where
the errors quoted were estimated from changes in $\chi^{2}$ values calculated using the confidence
ranges on $N_{H,i}$
produced from the spectral fits (and plotted in Fig. \ref{fig9}). We note that the parameters
which gave the minimum rms also gave the minimum $\chi^{2}$.
The rms scatter in the fitted $N_{H,i}$ values is now  $0.087$ dex, visibly less than for the distribution shown in Fig. \ref{fig8} and close to the 
value $\approx0.09$ dex expected from the original
distribution obtained from the individual spectral fits.
The reduced Chi-squared is now $\chi^{2}_{\nu}=11.1$ while the expected value
is $<\chi^{2}_{\nu}>=11.7$.
Including the molecular hydrogen column density
in the ISM model using the profile given by Equation \ref{eqH2} has eliminated the unexpected trend
in the fitted $N_{H,i}$ values seen in Figs. \ref{fig2} and \ref{fig5}.

In order to check that the results are not sensitive to the
particular distribution of GRBs within the 19 groups used, we performed the
analysis using different groupings. The mean and rms of
the fitted parameters taken across the groups are given in Table
\ref{tab1}. The mean photon indices and normalisations are reasonably constant.
When the number of GRBs in each group is only 13 the scatter in
the normalisations and excess column density values is largest, as
expected. For the grouping plotted in Fig. \ref{fig9}, $18\times26+1\times25$,
the scatter in excess column density is significantly lower.
Increasing the group size further to 29 makes little difference although
the scatter is now starting to increase. This is because some of the
groups now span too large a range in
Galactic absorption, a factor of 2 to 3,  and using a single,
fixed value for $N_{H,g}$ is no longer a good approximation for the
combined spectra. The results are clearly not sensitive to the particular
GRB groupings used although the chosen grouping gives the best
compromise between using too few GRBs per group and getting a large scatter
in $N_{H,i}$ and using too many GRBs per group and losing resolution in
$N_{H,g}$.

\begin{table}
\begin{tabular}{cccc}
Grouping & $N_{H,i}$ & $\Gamma$ & Norm \\
\hline
$37\times13+1\times12$ & $1.06\pm0.07$ & $1.87\pm0.16$ & $7.06\pm2.38$ \\
$25\times19+1\times18$ & $0.85\pm0.05$ & $1.84\pm0.13$  & $7.00\pm1.90$ \\
$18\times26+1\times25$ & $0.84\pm0.02$ & $1.83\pm0.12$ & $6.85\pm1.75$ \\
$17\times29$           & $0.82\pm0.03$ & $1.82\pm0.12$ & $6.90\pm1.78$ \\
\hline
\end{tabular}
\caption{The mean and rms scatter of fitted spectral parameter values 
using different groupings with $N_{H,g}$ from 
Equation \ref{eqH} and $N_{H_{2}}$ from Equation \ref{eqH2}.
The groupings are given as $\#$bins$\times\#$GRBs per bin.
$N_{H,i}$ 10$^{21}$ cm$^{-2}$ is the geometric mean of
the excess column density (at z=0).
The photon index is $\Gamma$ and the normalisation is Norm
10$^{-4}$ ph cm$^{-2}$ s$^{-1}$ keV$^{-1}$.}
\label{tab1}
\end{table}

We repeated this analysis using the trial1 and trial2  XRT calibrations
described in Section \ref{strial}. The best
fit parameter values were the same within the quoted errors. The
rms scatter was larger, $0.090$ dex using trial1 and $0.13$ using trial2.
The only really significant difference was the geometric mean of
$N_{H,i}$ across all the groups. In Fig. \ref{fig9} this is
$8.4\times10^{20}$ cm$^{-2}$ whereas using the trial1 calibration gave
$1.05\times10^{21}$ cm$^{-2}$ some 25\% higher and using trial2
gave $0.66\times10^{21}$ cm$^{-2}$, 22\% lower.

\section{Comparison of the $N_{H_{2}}$ distribution with other measurements}
\label{comparison}
\begin{figure}\begin{center}
\includegraphics[height=8cm,angle=-90]{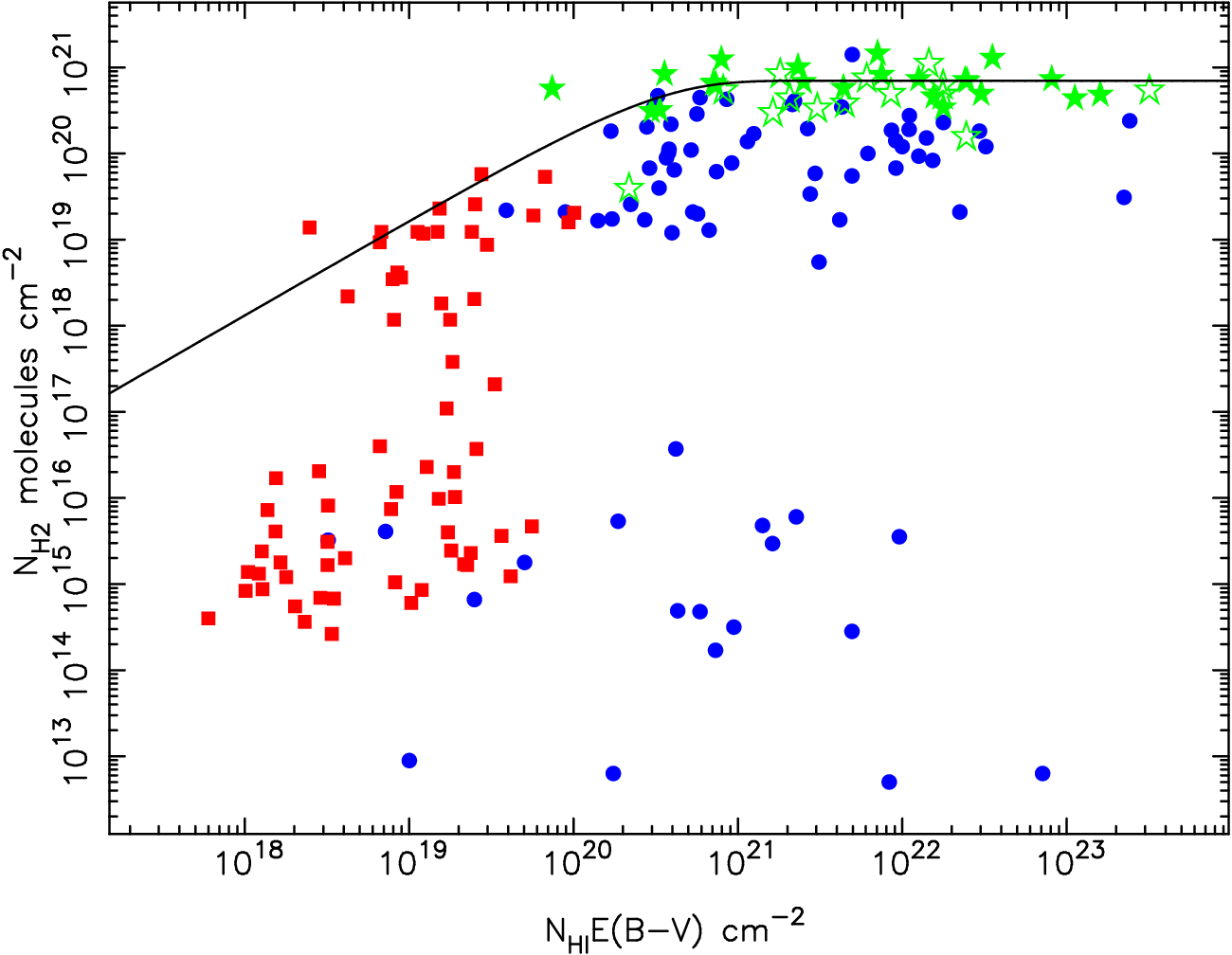}
\end{center}\caption{
The molecular hydrogen column as a function of
$N_{HI}E(B-V)$.
The {\em Copernicus}
measurements from Savage et al. (1977) are plotted as blue circles 
(sources with only upper limits for $N_{H_2}$ are not included).
The {\em FUSE} measurements from Rachford et al. (2002,2009)
are plotted as green stars/open stars and the
extragalactic {\em FUSE} measurements from Wakker (2006) are plotted as 
red squares.
The solid line is the profile inferred from the X-ray GRB afterglow spectra.}
\label{fig10}
\end{figure}

How does the molecular hydrogen column density profile inferred from GRB afterglow X-ray
spectra compare with other measurements? Fig. \ref{fig10} shows the molecular
hydrogen column measured in UV absorption to stars taken from 
\cite{1977ApJ...216..291S} (74 objects),
\cite{2002ApJ...577..221R,2009ApJS..180..125R} (38 objects)
and 66 extragalactic high latitude sources taken from
\cite{2006ApJS..163..282W}. The molecular hydrogen column density is plotted as a function 
of $N_{HI}E(B-V)$ derived from the LAB survey and 
{\em IRAS} and {\em COBE/DIRBE} extinction maps in the same way
as described above for implementing the $N_{H_{2}}$ profile 
(Equation \ref{eqH2}).
The profile derived from the X-ray absorption in GRB afterglows forms a reasonably well defined upper envelope to the distribution of $N_{H_2}$ points.
The plateau for $N_{HI}E(B-V)>N_{c}$ is well matched between the UV and X-ray 
confirming that $N_{H_{2}}$ is
independent of the atomic column density and the dust extinction in the Galactic plane.
There is clearly a very large range of UV measured values below the upper envelope. The X-ray
absorption measurements suggest that a large fraction of the molecular hydrogen
is not seen in the UV absorption spectra. This may be because the absorption line features seen 
in the UV are produced by dense clouds of gas which have a well defined velocity with little
dispersion. For many lines of sight through the galaxy a significant fraction of the molecular hydrogen
may be distributed at much lower density and with significant velocity dispersion. This material
would produce very broad absorption features in the UV spectra or possibly a large
number of small line features at different velocities which cannot be detected or identified.
Most of the {\em FUSE} observations of bright sources reported by
\cite{2002ApJ...577..221R,2009ApJS..180..125R} are close to
the X-ray profile indicating that when the signal to noise and sensitivity
are good all the molecular
hydrogen can be detected in UV absorption. It is possible that some of these
bright {\em FUSE} sources are not subject to the entire Galactic
column density. This might explain why
two sources, HD 164740 and HD 186994, have $N_{H_2}$ values
significantly lower than predicted by the model.

The {\em FUSE} source which is closest to a GRB position is
HD154368, 0.86 degrees from GRB 100504A. For this GRB the
$N_{H,g}$ is $2.6\times10^{21}$ cm$^{-2}$ and the fitted $N_{H,i}$ is 
$3.8\times10^{21}$ cm$^{-2}$. This is consistent
with the {\em FUSE} result of $N_{H_2}=1.45\times10^{21}$ cm$^{-2}$.
The next closest is HD40893, 1.64 degrees from GRB 061019 In this case
$N_{H,g}=3.8\times10^{21}$ cm$^{-2}$ and the fitted $N_{H,i}$ is
$8.4\times10^{21}$ cm$^{-2}$ consistent with the
{\em FUSE} value of $N_{H_2}=3.80\times10^{20}$ cm$^{-2}$.
{\em FUSE} source HD206267 is 1.71 degrees from GRB 050422
and {\em FUSE} source HD179406 is 1.77 degrees from GRB 060105 but these 
separations are now significant compared with the resolution
and variations in the $N_{HI}$ and $E(B-V)$ maps.
All other {\em FUSE} sources are several degrees or more from the nearest
GRB and any stacking analysis to try and make a more detailed comparison
is not possible.

Emission from the lower frequency rotational transitions of carbon monoxide (CO) molecules in the
radio band (the J=$1\rightarrow 0$ line is at 115 GHz)  is used as a proxy for $H_{2}$ in studies of
Giant Molecular Clouds,  (see e.g. \cite{1978ApJS...37..407D} and
\cite{2011MNRAS.412..337G}).
 A simple conversion factor 
$X_{CO}=N_{H_{2}}/W_{CO}\simeq 2\times 10^{20}$ cm$^{-2}$ K$^{-1}$ km$^{-1}$ s,
where $W_{CO}$ K km s$^{-1}$ is the intensity of the CO emission line (J=$1\rightarrow 0$)
integrated over velocity, is used to predict the $N_{H_{2}}$ column density of a cloud 
from the observed $W_{CO}$.
The density of the absorbing column can be estimated independently by various
means: from the diffuse $\gamma$-ray flux
produced by interactions with cosmic rays, from the measured extinction or from
the atomic hydrogen 21 cm emission, but the presence of $H_{2}$ is inferred and not observed directly.

\begin{figure}\begin{center}
\includegraphics[height=10cm,angle=0]{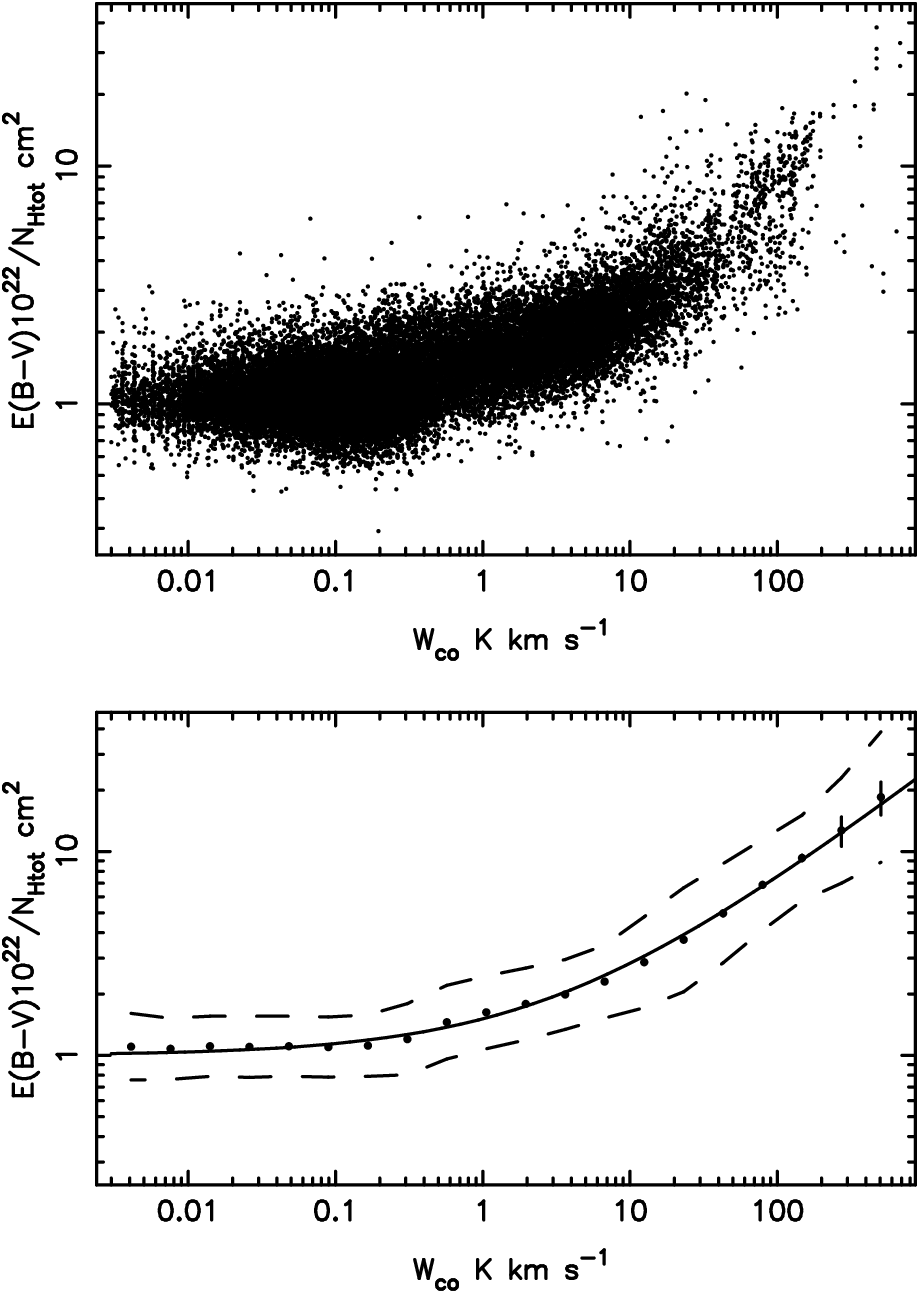}
\end{center}\caption{The dust-to-hydrogen ratio, $R_{DH}$, as a function of CO emission.
Top panel: Each point represents a pixel $\approx 0.6$ deg$^{2}$. $W_{CO}$ values are
available and plotted for $\sim60\%$ of the sky. Bottom panel: Average $R_{DH}$ values calculated
for 20 logarithmic bins across the full range of $W_{CO}$. The dashed lines indicate the 90\% range
of the scatter in $R_{DH}$. Error bars are plotted for the averages but these are too small to see except for
the bins at the top end. The solid line is the model fit described in the text.}
\label{fig11}
\end{figure}

 We attempted to
use such a scaling by rebinning the CO survey data from
 \cite{2001ApJ...547..792D}\footnote{Thomas Dame kindly provided us with the mid-latitude extension
of the CfA CO survey prior to publication and this gave us better coverage  out of the Galactic Plane.} to produce a CO map in the same Aitoff projection used by the FTOOLS NH procedure (as we did for
the $E(B-V)$ mapping described above). We extracted $W_{CO}$ values for every GRB position
and used these to set $N_{H,g}=N_{HI}+2X_{CO}W_{CO}$ in the fitting of the group spectra.
The extended CO survey covers $\sim60\%$ of the sky so we assumed that $W_{CO}=0$ for GRBs
which fell outside this area. This is reasonable because those sky areas
not covered by the survey are known to be regions of very low
CO emission.
We were unable to find a scaling factor $X_{CO}$ which reduced the rms scatter across the 19 groups.
In retrospect this is not surprising because the profile of the
molecular fraction ($f(H_{2})$ defined by Equation \ref{eqfh2})
predicted using the CO maps is
very scattered and not peaked around $N_{Htot}\simeq2\times10^{21}$ cm$^{-2}$ as required.
In fact there is no correlation between  $N_{H_{2}}$ given by Equation \ref{eqH2} and $W_{CO}$.
CO emission may be a proxy for molecular hydrogen in dense molecular clouds but this does not seem to
extend to the ISM distribution across the Galaxy. We searched for other potential correlations between
$W_{CO}$ and the dust or hydrogen column densities. We found that by far the best correlation was
with the dust-to-hydrogen ratio defined by
\begin{equation}
R_{DH}=\frac{E(B-V)\times10^{22}}{N_{HI}+2N_{H_{2}}}
\end{equation}
The factor of $10^{22}$ is included so that the ratio has a value $\sim1$ out of the Galactic
plane and Equation \ref{eqH2} was used to calculate $N_{H_{2}}$.
This ratio is plotted as a function of $W_{CO}$ in Fig. \ref{fig11}.
For the bottom panel we
calculated the average ratio values in 20 bins across the full $W_{CO}$ range. These average
values increase monotonically with $W_{CO}$ (within the noise)
and are well fitted by the function
\begin{equation}
R_{DH}=1+A\times W_{CO}^{\gamma}
\label{eqrdh}
\end{equation}
where $A=0.51\pm0.04$ and $\gamma=0.55\pm0.03$. This function fit is plotted in Fig. \ref{fig11}.
The correlation is still present if the molecular hydrogen component is ignored but the scatter is
significantly reduced when the $N_{H_{2}}$ term is included.

The dust-to-hydrogen ratio is important in the context of the X-ray absorption model of the
Galaxy. As defined by Equation \ref{eqrdh} the ratio changes from 1.0 out of the Galactic plane to a peak of
$\sim20$ in the Galactic plane. This increase could be due to a change in metalicity and imply a gradient
of the abundances of $A_{Z}$ in Equation \ref{eqgas} or it could reflect a change in the fraction of
the higher metallicity
material condensed into dust thereby representing a change in the $\beta_{Z,i}$ values
in the same Equation. We have tested both these possibilities by introducing, in turn, an abundance gradient
and a gas depletion gradient into the absorption model used for the spectral fitting of the GRB groups.
We only varied these factors for the elements expected to be incorporated into the dust
(C,O,Na,Mg,Al,Si,S,Cl,Ca,Cr,Fe,Co,Ni for which the gas depletion can be set in the tbvarabs routine
in XSPEC).
The group spectral fits are sensitive to the abundances: a change of just
20\% in abundance over the
full hydrogen column density range (i.e. $\pm10\%$ about the average given
by \cite{2000ApJ...542..914W})
is sufficient to significantly degrade the quality of the best fit model
shown in Fig. \ref{fig9}. The opposite is true for the gas depletion factors. If they are set to
the extreme value of 1.0 (no dust) then the quality of the fit degrades
but the changes
are barely statistically significant. Using a value of 0.0 (all dust) introduces an even smaller degradation. This is because changing the gas depletion does not alter the amount of 
material in the column but  simply changes the shielding effect of the dust grains and this
effect is small.
We conclude that the change in $R_{DH}$ must be dominated by gas depletion and not
an abundance gradient although the X-ray absorption measurements from the GRB afterglows
are unable to confirm the range in gas depletion implied by $R_{DH}$ shown in Fig. \ref{fig11}.
As already demonstrated, the spectral fitting of the grouped GRB
afterglow spectra serves as a good calibration of the elemental abundances assumed.
A study of Galactic absorption using the afterglow spectra from GRBs,
described by \cite{ 2011A&A...533A..16W}, concludes
that the metalicity of a typical
Galactic line of sight is not consistent with the abundances given by
\cite{2000ApJ...542..914W}.
However, if the molecular hydrogen column density is included in the
absorption model, we have shown that these abundances are correct to within
$\sim\pm10\%$ and there is no evidence for a large abundance gradient
as a function of hydrogen column density.

In general as $W_{CO}$ increases so the dust extinction, $E(B-V)$, and hydrogen column density,
$N_{Htot}$, increase, but the dust-to-hydrogen ratio given by Equation \ref{eqrdh} leads to
a rather counter-intuitive result: if $W_{CO}>10$ K km s$^{-1}$ then $R_{DH}\simeq 0.5 W_{CO}^{0.55}$.
If we measure the dust extinction and CO emission then we can estimate the total hydrogen column
 density as $N_{Htot}\simeq 2\times10^{22} E(B-V)W_{CO}^{-0.55}$ cm$^{-2}$.

The well known correlation between the dust column density (given
by $E(B-V)$ or $A_{V}=R_{V}E(B-V)$) and the hydrogen column
(given by $N_{HI}$) is better represented using $N_{Htot}$, including the
molecular hydrogen component.  The $A_{V}-N_{Htot}$ relation has a fractional
rms scatter of $\sim26\%$, less 
than the $A_{V}-N_{HI}$ relation for which the fractional rms scatter is
$\sim33\%$.
Because the dust-to-hydrogen ratio, $R_{DH}$, increases with $N_{Htot}$ the 
relationship is not linear. If $E(B-V)<0.1$, and adopting
$R_{V}=3.1$ from \cite{ 2011A&A...533A..16W}, then we find the linear
relation $N_{Htot}=3.2\times10^{21}$ cm$^{-2}$ $A_{V}$ 
but for $E(B-V)>0.1$ the correlation flattens off as $R_{DH}$
increases. If we ignore the molecular hydrogen we find
$N_{HI}=2.0\times10^{21}$ cm$^{-2}$ $A_{V}^{0.86}$ a result very similar
to the non-linear relationship given by 
\cite{ 2011A&A...533A..16W}. However, this non-linearity is
introduced because of the missing molecular hydrogen component
and the change in $R_{DH}$ and is not a consequence of a
gradient in metalicity.

\section{The molecular hydrogen fraction}
\begin{figure}\begin{center}
\includegraphics[height=8cm,angle=-90]{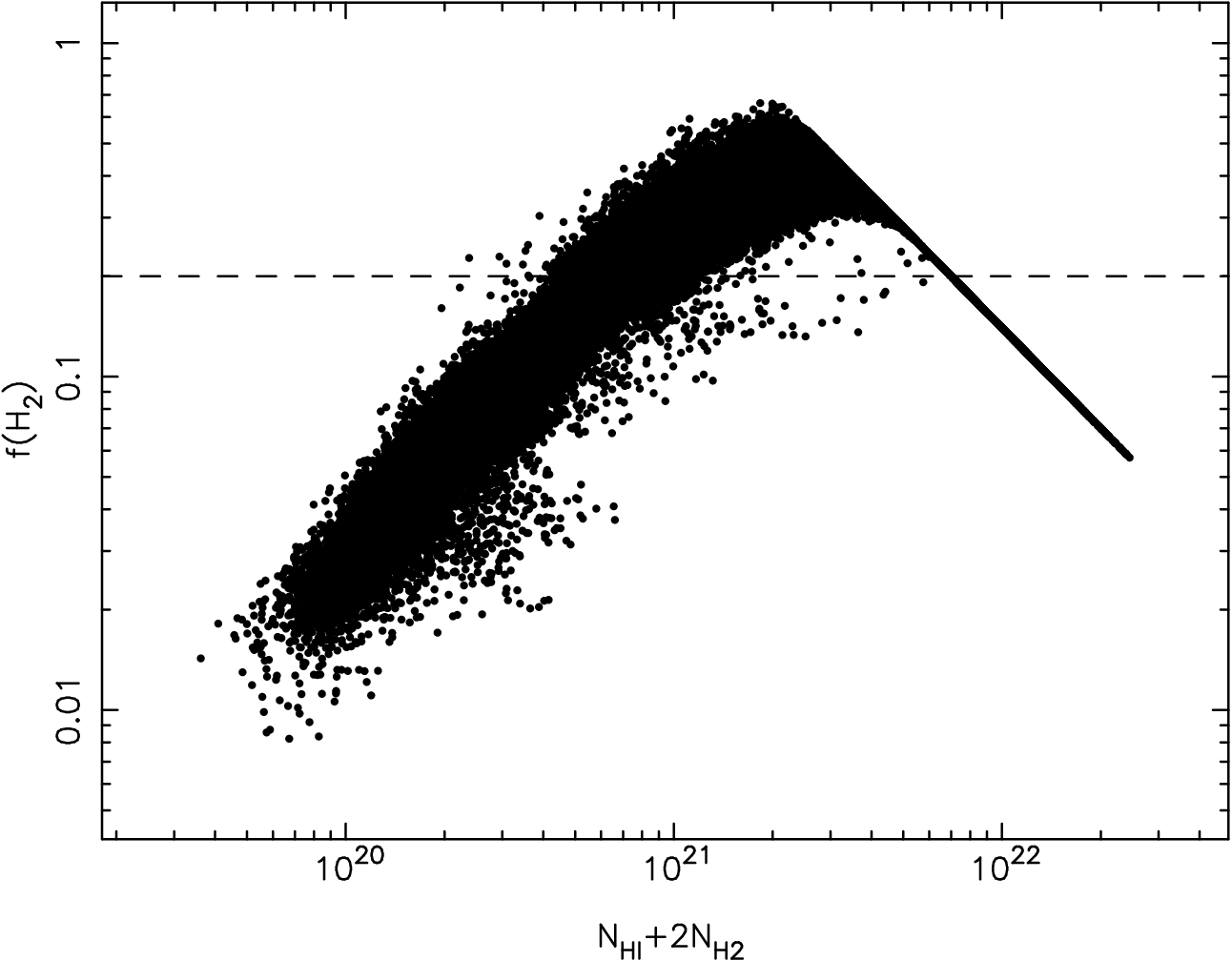}
\end{center}\caption{The molecular hydrogen fraction plotted as a function of the total
hydrogen column density calculated for pixels across the whole sky. The horizontal line represents the 20\% level
which is the default value in the XSPEC tbabs routine.}
\label{fig12}
\end{figure}

We can calculate the molecular hydrogen fraction using all the pixels in the FTOOLS version of the LAB survey all-sky image and the rebinned version of the E(B-V) data from \cite{1998ApJ...500..525S}. 
The result is plotted as a function of the
total hydrogen column density in Fig. \ref{fig12}.
Each point represents an estimate of molecular
hydrogen fraction for a sky area of $\approx 0.6$ deg$^{2}$.
The scatter is introduced because
there is considerable scatter in the correlation between $N_{HI}$ and $E(B-V)$.
At high column densities the
molecular hydrogen fraction actually decreases and the scatter gradually
disappears because
the $N_{H_{2}}$ column density asymptotes to the constant value,
$N_{H_{2}max}$.
Conversely, for low total hydrogen column densities
the molecular hydrogen density is increasing faster than
the total column density and we get a peak in the $f(H_{2})$ profile.
The presence of the peak in the $f(H_{2})$ distribution is why the introduction of
this model is able to improve the fitting of the GRB afterglow group spectra. It has the effect of
significantly increasing the absorption at and around $N_{HI}=10^{21}$ cm$^{-2}$ while having relatively
little impact on the  absorption at high and low $N_{HI}$ values. The model using the product $N_{HI}E(B-V)$ rather than
just $N_{HI}$  alone offers an improvement because the correlation between $N_{HI}$ and $E(B-V)$
is not tight and the molecular hydrogen column depends on both $N_{HI}$ and $E(B-V)$.
The constant value of 20\% used in the XSPEC tbabs routines is shown in Fig. \ref{fig12} as the 
horizontal dashed line whereas we find the maximum value of
$N_{H_2}$ is $\sim66\%$.
Remarkably the mean value of $f(H_{2})$ produced by the current model
over the sky is 20.2\%, very close
to the value adopted by \cite{2000ApJ...542..914W}.

Why does the molecular hydrogen fraction peak at
$N_{Htot}\approx2\times10^{-21}$
cm$^{-2}$
rather than continuing to rise as you might expect and why is the $N_{H_{2}}$
column density not correlated with the CO emission? 
We don't know, but we comment that
the formation and destruction of molecular species
is dependent on many factors. Theoretical and observational
studies of molecular clouds indicate that the formation of $H_{2}$ is
primarily determined by the time available for its formation
and the $X_{CO}$ factor
relating the presence of $H_{2}$ to the CO emission can
be highly variable if the metalicity and density are low and
the background UV radiation field is high
(\cite{2011MNRAS.412..337G,2011MNRAS.415.3253S}). The $N_{H_{2}}$ column density model
for the Galaxy presented here reflects the average state of the ISM of the Galaxy rather than
conditions in molecular clouds taken individually. There is a large scatter in both the
molecular hydrogen fraction and the dust-to-hydrogen ratio which must reflect a broad
spectrum of local conditions.

\section{Conclusion}
Measurements of X-ray absorption in the spectra of GRB afterglows 
indicate that using the distribution of $HI$ as the only direction
dependent variable in the model of the Galactic ISM is inadequate
to describe variations in the column density seen. The discrepancy
can be explained by including a molecular hydrogen column density
component which is a function of the product of the
$HI$ column density and dust extinction, $N_{HI}E(B-V)$, given
by Equation \ref{eqH2}. The
distribution of molecular hydrogen inferred from the X-ray absorption 
measurements is in agreement with direct measurements of the
$N_{H_{2}}$ column density made using UV absorption spectra.
Summing the column densities of atomic and molecular hydrogen we
can estimate the total hydrogen column density and calculate a 
dust-to-hydrogen ratio, $R_{DH}$. This ratio is shown to correlate with
the CO emission following the function given by Equation \ref{eqrdh}.
In principle $R_{DH}$ can be used to estimate the gas depletion
factors in the X-ray absorption model although the present
X-ray absorption measurements are not sensitive enough to warrant 
the inclusion of this direction dependence.

If the total effective hydrogen
column density (at $z=0$) is less than $\sim1.5\times10^{21}$
cm$^{-2}$ then the systematic error on the fitted value,
imposed by current uncertainties in the area calibration of the {\em Swift} XRT,
is $\sim16\%$. For higher column densities this systematic error
drops to $\sim8\%$. Furthermore, we can confirm
that the elemental abundances in the Galaxy
assumed in \cite{2000ApJ...542..914W} are correct to within $\sim \pm10\%$
although because of the limited spectral resolution of the CCD detector
the XRT is not
sensitive to the details of the ratios between the individual elements.

If you required an accurate estimate of the X-ray
absorbing column density in our Galaxy you should include
an estimate of the molecular hydrogen column
density\footnote{Software to
calculate $N_{H_{2}}$ and $N_{Htot}$ for a given position is available at
\url{http://www.star.le.ac.uk/~rw/xabs}}.
When using the tbabs in XSPEC you should, at the very least, multiply
the atomic hydrogen column density estimated from 21 cm survey data,
$N_{HI}$, by a factor of 1.25 so that
20\% of the total hydrogen column is molecular.
You can improve the accuracy by employing Equation \ref{eqH1},
calculating the total column using Equation \ref{eqH} and setting
the molecular fraction appropriately using the tbvarabs model.
If you have a Galactic $E(B-V)$ extinction measure for the source position
you can further improve the accuracy using the 
$N_{H_{2}}$ profile given in Equation \ref{eqH2}.

\section*{Acknowledgements}
We gratefully acknowledge funding for {\em Swift} at the
University of Leicester by the UK Space Agency and thank Thomas Dame for giving us
access to the extended CO survey data which have not yet been published. RLCS is supported by a Royal Society Dorothy Hodgkin Fellowship.

\bibliographystyle{mn2e}
\bibliography{rw_grbism}

\label{lastpage}
\end{document}